\documentclass[floatfix,prc,aps,twocolumn]{revtex4-1}
\usepackage{graphics,setspace,epsfig,color}
\usepackage{graphicx}
\usepackage{amscd}
\usepackage{amsmath}
\usepackage{amssymb}
\usepackage{amsthm}
\usepackage{float}
\usepackage{kantlipsum}
\usepackage{tabularx,ragged2e,booktabs}
\newcolumntype{C}[1]{>{\Centering}m{#1}}

\long\def\symbolfootnote[#1]#2{\begingroup%
\def\thefootnote{\fnsymbol{footnote}}\footnote[#1]{#2}\endgroup}

\newcommand{\be}{\begin{equation}}
\newcommand{\ee}{\end{equation}}
\newcommand\beq{\begin{eqnarray}}
\newcommand\eeq{\end{eqnarray}} 
\newcommand\nnnlo {N$^3$LO}
\newcommand{\E}

\usepackage{hyperref}

\begin{document}

\title{Microscopically constrained mean field models from chiral nuclear thermodynamics}

\author{Ermal Rrapaj$^{1,2}$}
\email{ermal@uw.edu}
\author{Alessandro Roggero$^2$}
\email{roggero@uw.edu}
\author{Jeremy W.\ Holt$^{1,3}$}
\affiliation{$^1$Department of Physics, University of Washington, Seattle, WA}
\affiliation{$^2$Institute for Nuclear Theory, University of Washington, Seattle, WA}
\affiliation{$^3$Cyclotron Institute, Texas A\&M University, College Station, TX}

\begin{abstract}

We explore the use of mean field models to approximate microscopic nuclear equations of state derived 
from chiral effective field theory across the densities and temperatures
relevant for simulating astrophysical phenomena
such as core-collapse supernovae and binary neutron star mergers.
We consider both relativistic mean field theory with scalar and vector meson 
exchange as well as energy density functionals based on Skyrme phenomenology and 
compare to thermodynamic equations of state derived from chiral two- and three-nucleon 
forces in many-body perturbation theory. Quantum Monte Carlo simulations of symmetric 
nuclear matter and pure neutron matter are used to determine the density regimes in which 
perturbation theory with chiral nuclear forces is valid. Within the theoretical uncertainties 
associated with the many-body methods, we find that select
mean field models describe well microscopic nuclear thermodynamics. 
As an additional consistency requirement, we study as well the single-particle properties of
nucleons in a hot/dense environment, which affect e.g., charged-current weak reactions in 
neutron-rich matter. The identified mean field models can be used across a larger range of
densities and temperatures in astrophysical simulations than more computationally 
expensive microscopic models.

\end{abstract}
\maketitle

\section{Introduction}
\label{intro}

Astrophysical phenomena such as core collapse supernovae and binary neutron star mergers, 
including the possible accompanying gravitational wave production and heavy-element 
nucleosynthesis in the matter outflow, are sensitive to the thermodynamics of isospin-asymmetric 
nuclear matter across many orders of magnitude in the nuclear density. In 
contrast to the hot ambient conditions characterizing these astrophysical environments, the 
structure and evolution of neutron stars are largely governed by the properties of cold and dense 
neutron-rich matter. Efforts are underway to combine astrophysical observations of neutron 
star properties \cite{LattimerPrakashScience,SteinerLattimerBrown,OezelReview2012,Demorest2010,Antoniadis2013}, terrestrial laboratory 
experiments of finite nuclei and hot/compressed nuclear matter \cite{Danielewicz02,Dutra2012}, and 
microscopic many-body theory \cite{Hebeler2010A,GCR2012,Krueger13,Roggero2014a,Wlazlowski14,Tews15} to construct 
a more complete picture of the nuclear equation of state (EOS) across the range of conditions probed 
in supernovae and neutron star merger simulations. 

Traditionally mean field models, based on either Skyrme phenomenology \cite{Lattimer91} or relativistic 
mean field (RMF) theory \cite{Shen91,Hempel12}, have been the method of choice for constructing nuclear
equations of state for numerical astrophysics simulations. The energy scales probed are much below the
regime where a perturbative treatment based on QCD, the fundamental theory of 
strong interactions, is feasible. Non-perturbative lattice QCD simulations may in principle overcome 
such difficulty, but at present multi-nucleon simulations are beyond the reach of modern supercomputers, 
although extracting a nucleon-nucleon interaction from lattice data is a topic of intense current research 
(see e.g., \cite{Aoki2011,Aoki2012,Savage2012,Takashi2013}). An alternative is to construct a low-energy
theory of strongly interacting matter, chiral effective field theory \cite{Weinberg79,Entem03,Epelbaum09,Machleidt11},
which has already had many successes in the description of nuclear structure and reactions of light 
and medium-mass nuclei \cite{Kalantar2012,Barrett2013,Roth2011,Hergert2013,Epelbaum2010, Hagen2012A,Otsuka2010,Holt2012,Soma2013, Wienholtz2013,Hagen2015},
as well as nucleonic matter at zero temperature \cite{Kaiser2002,Holt11,Holt2013b, Gezerlis2013,Coraggio13,Tews2013,Baardsen2013,Hagen2014,Roggero2014a,Roggero2014b}.
In recent years nuclear thermodynamics \cite{Tolos08,Fiorilla12,Holt13,Wellenhofer2014,Wellenhofer2015} from 
chiral effective field theory has been successful
at describing the homogeneous phases of symmetric and asymmetric matter (including the equilibrium mixed 
liquid-gas phase) up to densities $\rho \simeq 2\rho_0$, where $\rho_0 = 0.16$\,fm$^{-3}$ is the saturation 
density of nuclear matter, and temperatures up to $T \simeq 25$\,MeV. However, the construction of an 
equation of state for direct use in astrophysical simulations is still in progress, since the description of 
nuclear matter at larger densities and temperatures is currently outside the scope of many-body 
perturbation theory with coarse-resolution chiral nuclear potentials. Here phenomenological
mean field models can be used to extend the description, provided they match onto 
the low-energy theory.

In this work we analyze both Skyrme and RMF models for consistency with chiral nuclear thermodynamics. 
That is, we explore what conditions are required for mean field models \cite{BrownAchim,Davesne15,Bulgac15}
 fit to
the zero-temperature equation of state to be also consistent
with the finite-temperature equation of state and transport properties of nuclear matter.
In particular we will show that it is possible to 
find both (i) non-relativistic (i.e., Skyrme-type) as well as (ii) relativistic mean field models (RMF) 
that produce a thermodynamic equation of state for both 
pure neutron matter (PNM) and symmetric nuclear matter (SNM) compatible with many-body 
perturbation theory (MBPT) 
calculations employing chiral two- and three-nucleon interactions. Such models are also compatible with 
recent experimental constraints on nuclear matter properties \cite{Dutra2012}, and their use in astrophysical
simulations may therefore serve as an accurate substitute for the more computationally demanding 
microscopic chiral equations of state.

%
%
The manuscript is organized as follows.
In section \ref{benchmark} we compare MBPT calculations at both second- and third-order to 
non-perturbative quantum Monte Carlo calculations at zero temperature to assess uncertainties 
coming from neglected correlations in the perturbative treatment. This analysis is used in 
Section \ref{MFEOS} to identify mean-field models for the zero-temperature equation of state compatible 
with results obtained from a coarse-resolution ($\Lambda = 414$\,MeV) chiral interaction 
at next-to-next-to-next-to-leading order (N$^3$LO) \cite{Coraggio2014}. 
We then analyze the consistency of mean field models with chiral nuclear thermodynamics in section 
\ref{thermodynamics} as well as nuclear single-particle properties in Section \ref{differences} that affect
charged-current weak reactions in proto-neutron stars. We end with a summary and conclusions.  

\section{Ab initio uncertainty estimates on microscopic EOS models}
\label{benchmark}

In the present section we analyze the theoretical uncertainties from many-body 
perturbation theory in computing the ground state energy of symmetric nuclear matter 
and neutron matter. As a nonperturbative benchmark we employ quantum Monte Carlo techniques.
The goal is to identify the nuclear force models and density regimes where microscopic many-body 
calculations can impose the strictest constraints on phenomenological mean field models.
One may already anticipate that interactions
with low-momentum regulators would perform qualitatively better in this regard, especially in neutron matter,
however our goal here is to have a quantitative understanding of these errors for a relevant set of potentials.

\subsection{Method}
\label{smethod}

The quantum Monte Carlo calculations presented here are based on the recently developed 
configuration interaction Monte Carlo (CIMC) 
method~\cite{Mukherjee2013,Roggero2013,Roggero2014a}. Projection QMC methods like CIMC are based
on filtering out an eigenstate $\lvert\Psi_0\rangle$ of the Hamiltonian $H$  by
repeated application of the propagator $\mathcal{P}=e^{-\Delta \tau(H-E_T)}$ on an initial state 
$\lvert\Psi_{\rm I}\rangle$:
\begin{equation}
\lvert\Psi_0 \rangle = \lim_{N_{\tau} \to \infty} \mathcal{P}^{N_{\tau}} \lvert \Psi_{\rm I} \rangle . 
\end{equation}
Here, $N_\tau$ is the number of imaginary time steps, $E_T$ is an energy shift used to keep the 
norm of the wave function approximately constant, and $\Delta \tau$ is a finite 
step in `imaginary' time: $\Delta \tau=i\Delta t$. The state $\lvert\Psi_0\rangle$ is the eigenstate with the lowest eigenvalue 
within the subset of states having 
non-zero overlaps with $\lvert \Psi_{\rm I} \rangle$. The application of the propagator is carried out stochastically.

The main difference between the CIMC method and traditional continuum projection Monte Carlo methods
is that in the CIMC method this stochastic projection is performed in Fock space (i.e., the basis is provided by the 
Slater determinants that can be
constructed from a finite set of single particle (sp) basis states), as 
opposed to in coordinate space. As a result, non-local Hamiltonians do not pose any technical 
problems. In this work, we use the sp basis given by eigenstates of momentum and the $z$ components of spin and isospin.
A finite sp basis is chosen by imposing a ``basis cutoff" $k_{\rm max}$ and convergence is checked by performing
a sequence of calculations with increasingly larger values of the cutoff. This same sp basis is then employed to
perform many-body perturbation theory calculations at both $2$nd and $3$rd order (see e.g., Ref.\ \cite{ShavittBartlett}).

In order for a stochastic sampling to be feasible, the matrix elements of the propagator, 
$\mathcal{P}$, need to be positive semi-definite. For nucleons interacting with realistic interactions this condition 
is never fulfilled due to the presence of repulsive contributions, giving rise to the so-called sign problem. In CIMC we circumvent this by using a coupled-cluster 
double (CCD) wavefunction to restrict the random 
walk in a subsector of the full many-body Hilbert space where the positivity of the propagator is guaranteed 
(for details see Appendix~\ref{ssignproblem} and Ref.\ \cite{Roggero2013}). 
This introduces a systematic bias in the calculation (see Appendix~\ref{ssignproblem} for an attempt to estimate 
its effects), but nevertheless the final energy eigenvalue $E_0^{FN}$ is guaranteed to be an upper bound of 
the true ground state energy $E_0$.

The formalism is fully compatible with the use of a three-body interaction in the hamiltonian (eg. in the 
same way as these are dealt with in CC-theory \cite{Hagen2014}), however in order to 
have more control on the fixed-node procedure the explicit inclusion of triplet correlations in the coupled cluster
wavefunction has to be accounted for. Presently this poses strong limitations on the size of the sp basis that
can be handled, making it difficult to reach convergence with respect to $k_{\rm max}$. An alternative approach
based on cancellation techniques (see e.g., Ref.~\cite{Booth,Petruzielo2012,Kolodrubetz2012}), which may allow one
to neglect explicit three-body correlations, is currently being explored.

\subsection{Results}

In this study of the convergence of MBPT we will use four different chiral NN interactions: the 
NNLO$_{opt}$ from Ekstr{\"o}m et al.~\cite{Ekstrom2013}, the two low-momentum cutoff interactions 
from Coraggio et al.\ \cite{Coraggio2014} at N$^3$LO with cutoffs of $414$ MeV
and $450$ MeV and finally a version of the N$^3$LO interaction with a $500$ MeV cutoff from Entem and 
Machleidt \cite{Entem03,Machleidt11} where the N$^3$LO contact terms have been refitted using 
the same regulator for all partial waves.

Calculations at constant density are performed by first choosing the number of particles in the system $N$ 
and the target density $\rho$. The size $L$ of the box is then given by $L = (N/\rho)^{1/3}$. In the present 
simulations the number of particles is fixed at the closed shell $N=14$, and all of the results 
are converged with respect to the single-particle momentum cutoff $k_{\rm max}$, and we 
find that values of $k_{\rm max}$ slightly higher than the cutoff $\Lambda$ in the regulating function of
the chiral potentials is needed 
for convergence (this is in good agreement with the findings in Ref.\ \cite{Roggero2014a}).
We have carefully examined the dependence on the system size and found that $N=14$ provides reliable
estimates at the low densities we are interested in, a more detailed discussion is provided in Appendix~\ref{fsize}. 
Furthermore, all the perturbative calculations are performed by including corrections to the nucleon self-energy 
at the Hartree-Fock level, in other words by performing normal ordering with respect to the Hartree-Fock state 
(for additional details see e.g., Ref.\ \cite{ShavittBartlett}).
More detailed comparisons to results obtained using a free-particle spectrum can be 
found in Appendix~\ref{sspspectrum}.

\subsubsection{Pure Neutron Matter EOS}

\begin{figure}[t]
\begin{center}
\includegraphics[scale=0.35]{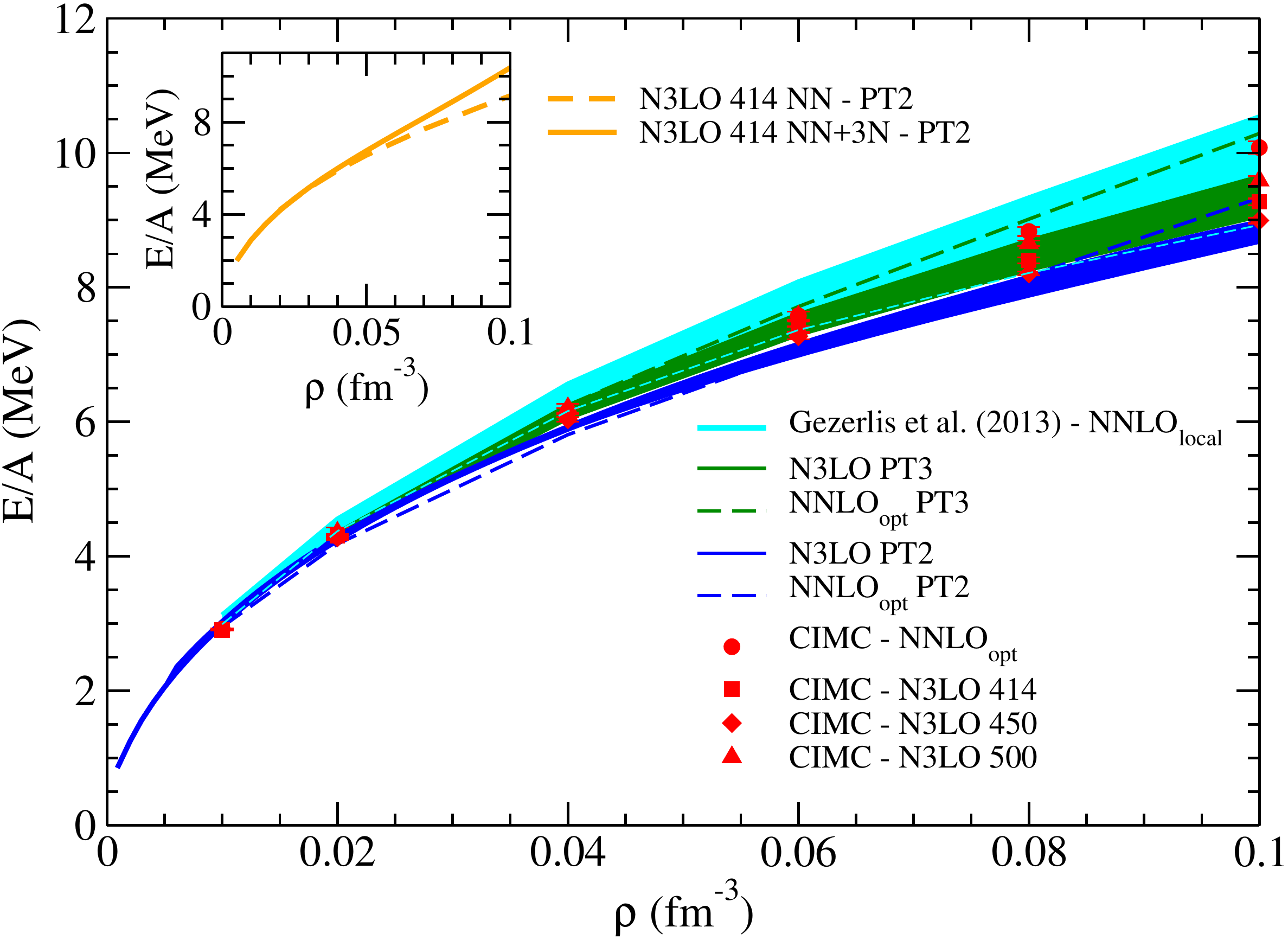}
\caption{(color online) Energy per particle of pure neutron matter as a function of density 
from different chiral interactions at NNLO and N$^3$LO. Results at second-order (PT2) and 
third-order (PT3) in perturbation theory are shown together with those from configuration 
interaction Monte Carlo (CIMC) and auxiliary field diffusion Monte Carlo
\cite{Gezerlis2013}. The inset shows the results at second-order in MBPT for the \nnnlo~$414$ 
interaction with and without three-nucleon forces.}
\label{Fig_nmat_rel}
\end{center}
\end{figure}

We first concentrate on the equation of state of pure neutron matter, shown in Fig.\ \ref{Fig_nmat_rel},
for both perturbative methods and the nonperturbative CIMC method employing chiral two-body forces.
The first qualitative observation that can be made is that the NNLO result lies consistently outside 
the \nnnlo~band obtained by cutoff variation. However, as noted in Refs.\
\cite{Krueger13,Epelbaum2015,Sammarruca15} this variation may not be a reliable estimate of the spread 
in predictions at a given order in the chiral expansion.
From the inset in Fig.\ \ref{Fig_nmat_rel} we observe that three-body
interactions are needed for a quantitative study for densities $\rho \gtrsim 0.06$ fm$^{-3}$. In the present section
we focus on NN forces only, and consequently this is the highest density where we may draw rigorous conclusions.

\begin{figure}[t]
\begin{center}
\includegraphics[scale=0.35]{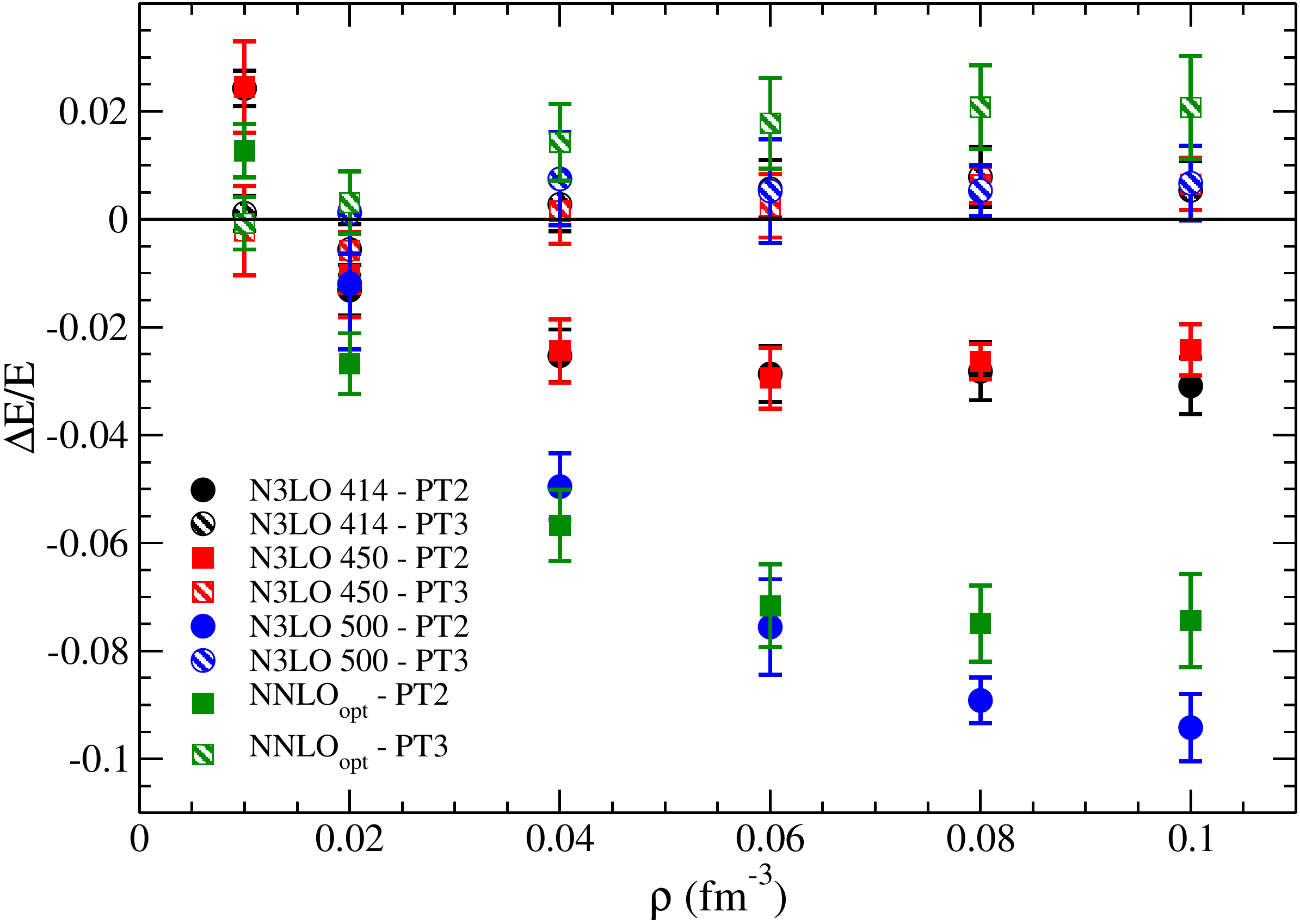}
\caption{(color online) Relative differences in energies obtained at different orders in perturbation theory
with respect to the corresponding CIMC result for pure neutron matter. Notation same as in Fig.\ 
\ref{Fig_nmat_rel}.} 
\label{Fig_nmat_evscimc}
\end{center}
\end{figure}

Evaluating the convergence of the perturbative calculations in Fig.\ \ref{Fig_nmat_rel}, we see that in
PNM the third-order contributions are generally sufficient to bring the MBPT results very close to those from 
the non-perturbative CIMC, indicating a good convergence pattern for MBPT for all four interactions considered, 
regardless of the regularization cutoff.
To estimate the errors corresponding to the perturbative convergence, we consider in Fig.\ 
\ref{Fig_nmat_evscimc} the relative difference between the CIMC results and the energies obtained 
at different orders in perturbation theory. We observe that the interactions split
into two groups with \nnnlo~$414$ and $450$ deviating from CIMC by $\approx 2-3 \%$ at second
order and the NNLO$_{opt}$ and \nnnlo~$500$ interactions that have deviations of $6-8 \%$. For
all interactions the third-order calculations are compatible with CIMC for all densities apart
from NNLO$_{opt}$, which at the highest densities has a small deviation of $\approx 1-2\%$.

One can study the effect of the bias coming from the fixed-node procedure by adding to the wavefunction a 
non-zero overlap with all states
in the Hilbert space (triplet and higher order irreducible correlations) and checking the differences in the 
final estimate for the energy. In the case of PNM and for all four interactions this procedure gives 
results compatible with the energies obtained by employing the CCD wavefunction, which indicates
that this source of systematic error is under control (a more detailed study of this systematic error is presented in Appendix~\ref{ssignproblem}).
 
Before concluding this section we comment on the low density regime of neutron matter, which due
to the large neutron-neutron scattering length ($a_{nn} \sim -19$\,fm) is very close to that of a
non-perturbative unitary fermi gas. From the results in Fig.\ \ref{Fig_nmat_evscimc}
we may draw the (wrong) conclusion that at low densities ( $\rho \lesssim 0.02$ fm$^{-3}$ ) the perturbative 
calculations converge faster, with the second-order results at $\rho\approx0.02$ fm$^{-3}$ almost compatible 
with CIMC. This is however an artifact coming from the fact that the second-order predictions change from 
too attractive at high density to too repulsive at low ones, that is, the relative difference with CIMC changes 
sign at low densities (as is evident from Fig.\ \ref{Fig_nmat_evscimc}). This is then in agreement with the 
expectation that at sufficiently low densities the convergence of MBPT breaks down.

\subsubsection{Symmetric Nuclear Matter (SNM)}
\begin{figure}[t]
\begin{center}
\includegraphics[scale=0.35]{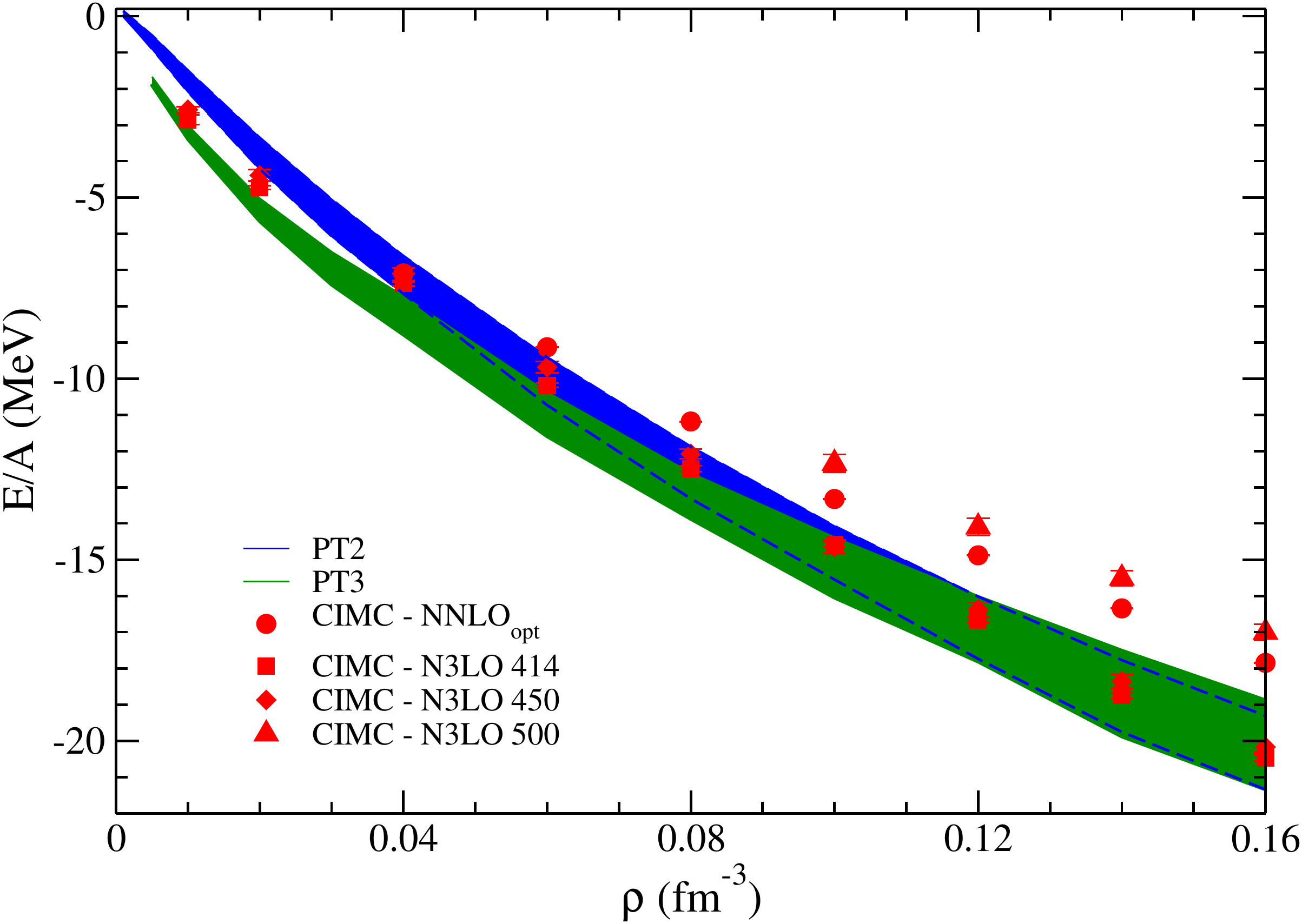}
\caption{(color online) Same as in Fig.~\ref{Fig_nmat_rel}, except for symmetric nuclear matter. 
The NNLO$_{opt}$ results now are contained in the \nnnlo~bands. } 
\label{Fig_smat_rel}
\end{center}
\end{figure}

\begin{figure}[t]
\begin{center}
\includegraphics[scale=0.35]{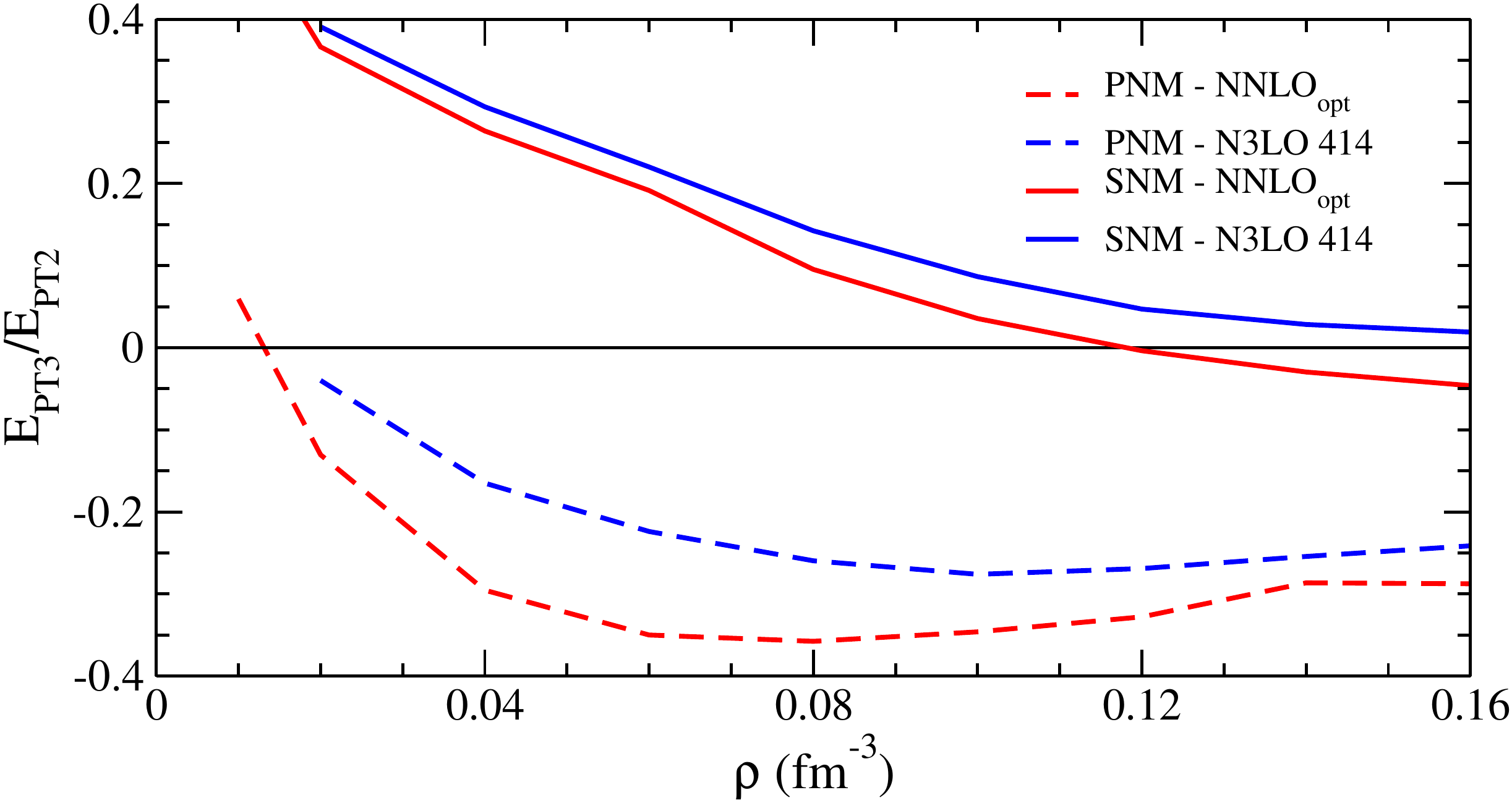}
\caption{(color online) Ratio of the third-order correction to the second-order one in PNM (dashed lines) and SNM (full lines) for 
NNLO$_{opt}$ (red) and \nnnlo~414 (blue).} 
\label{Fig_smat_rel_inset}
\end{center}
\end{figure}

We now focus on the results for symmetric nuclear matter, shown in Fig.\ \ref{Fig_smat_rel}, 
that have been computed using $A = 28$ nucleons. In contrast to the 
neutron matter equation of state in the last section, here the various chiral two-nucleon 
forces give rise to larger variations. As a result the energy per particle from the NNLO interaction 
is now contained in the \nnnlo~band, 
which is mostly due to the \nnnlo~$500$ results that are consistently less attractive than the rest
for intermediate to high densities. Reaching low densities with this
high-cutoff interaction is computationally more demanding, and since 
we already observed a lack of convergence in the simpler case of neutron 
matter, we will discuss in the following only the remaining three interactions. Another interesting feature 
apparent from the data (and 
compatible with earlier findings by Coraggio et al.\ \cite{Coraggio2014}) is that at intermediate to high
densities the third-order perturbative correction is negligible. To better understand this behavior we plot 
in Fig.\ \ref{Fig_smat_rel_inset} the ratio between third- and second-order perturbative corrections in both
neutron and symmetric nuclear matter. We see that
the third-order correction slowly changes sign around the empirical saturation density, giving an artificially small correction to the energy per particle.

If we were to evaluate the perturbativeness of the calculation by only comparing the magnitudes 
of the contributions at different orders in perturbation theory 
(see e.g., Fig.\ \ref{Fig_smat_rel_inset}), we might draw
the counterintuitive (and wrong) conclusion that the convergence pattern in SNM is much better than in PNM.
That this is not the case can be understood by noticing that if the third-order calculation is
already converged then the CCD wavefunction used in the fixed node procedure would be the exact
ground state, but if that were the case then the third-order perturbative results and the ones predicted by 
CIMC should be on top of each other, which is not the case. 

Another way to see this is to exploit the 
freedom in CIMC to tune the guiding wavefunction by truncating contributions beyond a given number of
particle-hole excitations. The second-order energies contain information about 2p-2h excitations
above the Hartree-Fock ground state, so any important effects coming from higher orders cannot
be recovered in this low truncation of the MBPT expansion. We tested the NNLO$_{opt}$ interaction 
at $\rho=0.08$ fm$^{-3}$ in both PNM and SNM and found that in the first case in order to recover
the full CIMC results we have to allow up to 6p-6h contributions in the guiding wavefunction
(with the 4p-4h truncation lying just outside the error bands) while for the latter case
up to 12p-12h states are needed to reproduce the non-truncated CIMC result. Similar observations hold for
the \nnnlo~414 interaction that is used in subsequent sections of the paper, with 4h-4p being converged in 
PNM and 10p-10h for SNM. This
observation is again in strong favor for enhanced non-perturbative features in the symmetric 
nuclear matter case, even for an interaction with a low-momentum cutoff. 

We conclude that it may be misleading to evaluate the convergence
of second-order calculations using only the magnitude of the next order as guidance 
(or equivalently using Pad\'e extrapolations), especially in regimes where the third-order 
contributions are changing sign and are thus artificially small. Without the fourth-order results 
it is then difficult to judge
convergence within MBPT itself and comparison with non-perturbative methods are important.

\begin{figure}[t]
\begin{center}
\includegraphics[scale=0.35]{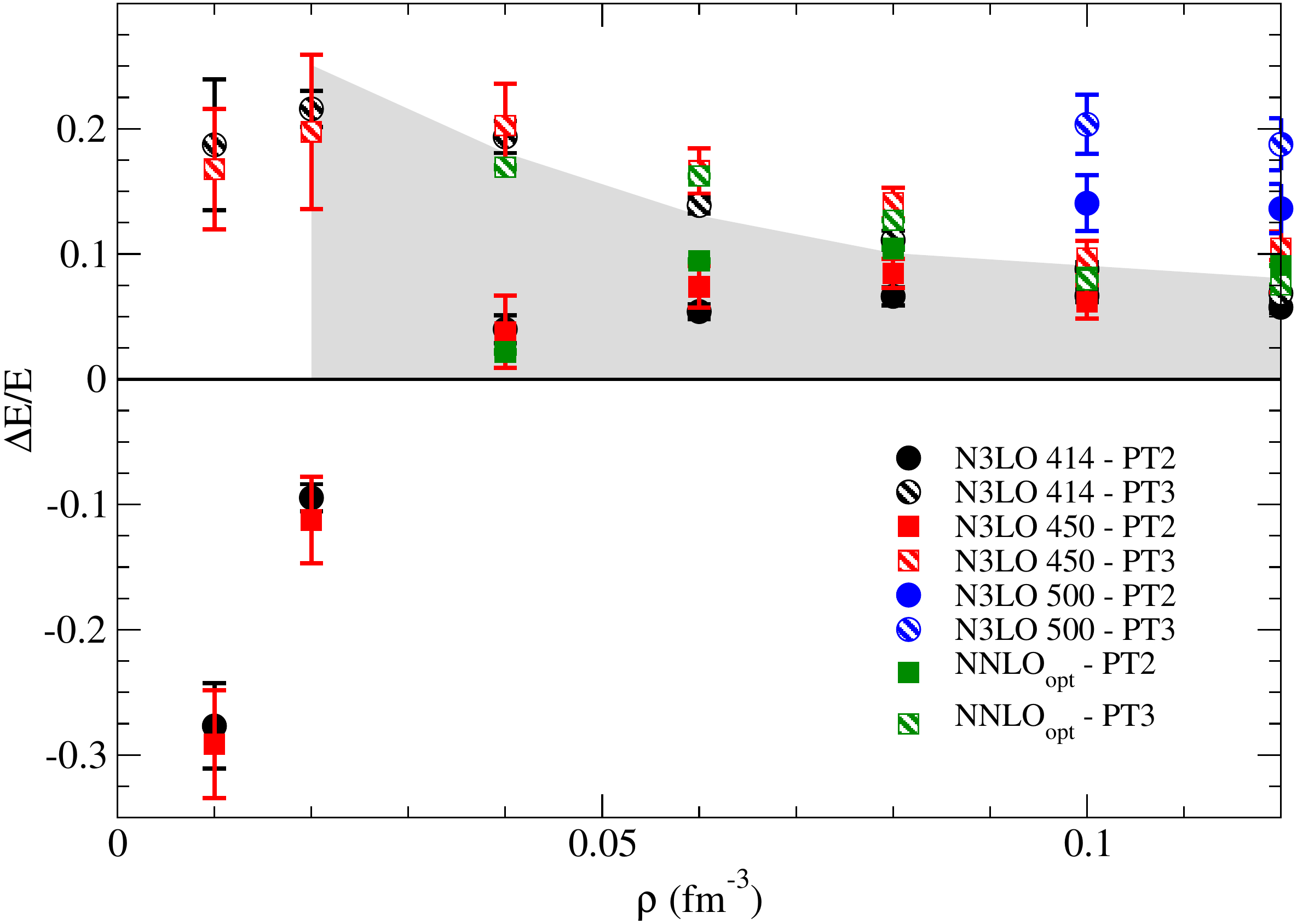}
\caption{(color online) Relative deviations of energies obtained at different orders in perturbation
theory with respect to the corresponding CIMC result for symmetric nuclear matter. The gray band 
is a qualitative estimate of the uncertainties in the fixed-node approximation (see text for the details).} 
\label{Fig_smat_evscimc}
\end{center}
\end{figure}

In Fig.\ \ref{Fig_smat_evscimc} we plot the difference between the second- and third-order 
perturbative results for the symmetric matter equation of state and that obtained from CIMC. 
Recall that the
main systematic error in the Monte Carlo calculations, the fixed-node error, is thought
to be much more severe in this system especially at low densities due to the appearance of
bound states that are extremely difficult to capture explicitly in the coupled-cluster ansatz
for the guiding wavefunction. One of the appealing properties of the CIMC method however
is that it is guaranteed to give an upper bound to the energy. As was mentioned 
above, by looking at the relative variations in Fig.\ \ref{Fig_smat_evscimc} we can get the impression
that second-order calculations provide as good (or better) estimates of the energy than those obtained at 
third order in the density range $0.04 - 0.10$ fm$^{-3}$. This is a well known feature of asymptotic expansions 
(see e.g., Ref.\ \cite{olver1974})
like the one employed in MBPT (which does not involve any small parameter), where one finds
in fact that the smallest error achievable (without employing resummations) is obtained by 
truncating the series at an optimal order $N^*$ which may 
indeed be very small. However, given the present large uncertainties, no conclusive statement can
be made. 

A more rigorous statement can be made instead for lower densities
$\rho < 0.04$ fm$^{-3}$ by using the upper bound nature of CIMC energies: in the low density 
regime the errors introduced by stopping the perturbative expansion at second order
are larger than $\approx 20\%$. One should be cautious about any statement on properties
of nuclear matter in this regime, such as the spinodal instability region, coming from second-order 
MBPT. In order to have more control in that region, comparisons to higher-order
results are therefore needed before a better understanding of non-perturbative physics is achieved.
In the intermediate-density regime, both second and third-order calculations of the ground state
energy are very similar to each other and on average $7-10 \%$ more bound than CIMC results. 
Due to the lack of control on the fixed-node errors, this would be the most reliable error band we 
can attribute to perturbation theory, but due to its qualitative nature it will not be used to discriminate 
among the various mean field models in subsequent sections. We note however that this estimate is 
consistent with the findings of Hagen et al.\ \cite{Hagen2014} in which the NNLO$_{opt}$ interaction 
resulted in a difference between CCD and MBPT at second order of $\approx \!12\%$ at 
$\rho=0.06$ fm$^{-3}$ to $\approx \! 8\%$ around saturation density.


\section{EoS from Mean Field Models }
\label{MFEOS}

In the previous section we found that the two coarse-resolution chiral interactions at \nnnlo~with cutoffs
$\Lambda =$ 414 and 450 MeV have good perturbative properties in PNM and SNM, with average errors 
on the order of $\Delta E / E \approx 2-3 \%$ for the former and $\Delta E / E \approx 8-10 \%$ for the latter in the 
vicinity of nuclear matter saturation 
density. In addition, as shown in fig.~\ref{Fig_nmat_rel}, for densities up to $0.05 - 0.06\ \text{fm}^{-3}$, the contribution to the
energy per particle from 3N forces is negligible and the error estimates from our analysis can be used quantitatively to select mean field models that 
are consistent with second-order MBPT calculations at low densities. In the following we will adopt the most perturbative \nnnlo~414 interaction including both NN and 3N forces.

In astrophysical simulations of core-collapse supernovae and neutron star mergers, equations of state 
based on mean field models have been used for more than three decades.
They provide a practical way of describing the physical properties of dense matter while being 
computationally inexpensive, thus
allowing for the fast generation of data needed for simulations. Among these models those derived from 
Skyrme interactions are widely used for the description of both nuclei and infinite matter, with a prominent 
example being that of Lattimer and Swesty \cite{Lattimer91}. 

The general expression for the energy density functional of homogeneous matter is given by
\begin{eqnarray} \nonumber
 \mathcal{E}&=& \sum_{q=n,p}\frac{1}{2M_q}\tau_q+\frac{1}{4}t_0\Big[(2+x_0)\rho^2 \\ \nonumber
 &-& (1+2x_0)\sum_{q=n,p}\rho^2_q\Big]
 +\frac{1}{8}\Big[ a \tau \rho+2b\sum_{q=n,p}\rho_q \tau_q\Big]\\ 
&+&\frac{1}{24}t_3\rho^{\alpha}\Big[(2+x_3)\rho^2-(1+2x_3)\sum_{q=n,p}\rho_q^2\Big],
\label{eq:SkyrmeEDF}
\end{eqnarray}
where $\rho$ is the density and $\tau$ is the kinetic energy density. Skyrme interactions represent
a low-momentum expansion of an effective two-body NN interaction and are usually constrained by the
properties of finite nuclei close to the valley of stability. 
The predictive power of Skyrme models however is difficult 
to assess away from the region in parameter space where it has been fitted.
The astrophysical environments mentioned above span a wide range of densities, from a rarefied gas 
to dense matter possibly 2-5 times that of normal nuclei, and neutrons represent the 
vast majority of the nuclear composition. Moreover, these environments are characterized by high 
temperatures that can reach up to $T = 50$\,MeV \cite{Janka2015}. The reliability of effective Skyrme interactions
to accurately cover this range of parameter space based on fits to terrestrial nuclei is an open question.

A first consistency check can be carried out using experimentally extracted constraints for nuclear
matter properties. In Ref.\ \cite{Dutra2012} Dutra et al.\ have benchmarked a set of $240$ Skyrme 
parametrizations against a set of $11$ infinite matter constraints. 
Only $16$ parametrizations were shown to meet all of these constraints. Due to the wide 
range of conditions encountered in supernovae simulations, a careful calibration of the models
also in conditions that are difficult to achieve in terrestrial experiments is therefore important. Recent 
developments in both many-body techniques and nuclear forces allow
the equation of state of PNM to be relatively well constrained at low densities. Using these results, 
together with properties of
double-magic nuclei \cite{Brown2013}, Brown and Schwenk \cite{BrownAchim} have refitted from 
the consistent models in Ref.\ \cite{Dutra2012} a selected smaller 
set of six ``best fit" Skyrme models: SKT1, SKT2, SKT3, SKa25s20, Ska35s20, Sv-sym32. In 
Ref.\ \cite{BrownAchim} two slightly different parameter sets for these interactions have
been obtained by requiring a particular value for the effective mass ($m^*/m = 0.9$ or $1.0$) in neutron matter 
at the density $\rho = 0.10$ fm$^{-3}$. 

In this work we will consider
all of these final $12$ parametrizations. For mean field Skyrme models, theoretical uncertainties
can be estimated through a careful extraction of the correlation matrix for the
parameters entering the functionals \cite{Klupfel2009,Dobaczewski14,Kortelainen2015}. 
In recent years important efforts have been devoted by the UNEDF
collaboration to construct Skyrme parametrizations with reliable error quantifications; for this reason 
we have also considered the parametrizations UNEDF0 \cite{UNEDF0_ref}, 
UNEDF1 \cite{UNEDF1_ref} and UNEDF2 \cite{UNEDF2_ref}. For the other 
parametrizations there are no publicly available parameter correlations that
can be used to study uncertainties and propagation of systematic errors. It is important to
realize however that detailed error extrapolations in complicated simulations (like those for 
core--collapse supernovae) poses major technical challanges and therefore parametrizations whose
central values are already close to the expected ones are to be preferred. In this work, we therefore
focus on constraining mean field models from the low-density microscopic equation of state and study 
their properties at saturation density as described by Table \ref{table:constraints} assuming
negligible errors in the parameters. However, we would like to stress the need for a detailed 
study of parametric uncertainties, which is outside the scope of this study. Such analysis would
be greatly facilitated if new parametrizations would be published together with their parameter's 
covariance matrix informations.

Recent astrophysical simulations also employ relativistic mean field models of the nuclear equation of state.
The general expression for the energy density functional in terms of nucleon $\Psi$, scalar $\phi$, vector $V_\mu$,
and vector isovector $b_\mu$ fields is given by
\begin{eqnarray} \nonumber
\mathcal{E} &=& \sum_{s,t}\int\frac{d^3k}{(2 \pi)^3}\sqrt{k^2+(M_{t}-g_S \phi_0)^2}(f_{s,t}+\overline{f}_{s,t})\\ \nonumber
&+&\frac{1}{2}m_s^2 \phi_0^2+\frac{\kappa}{3!}(g_S \phi_0)^3 +\frac{\lambda}{4!} (g_S \phi_0)^4+\frac{\zeta}{8} (g_V V_0)^4\\
&+&\frac{1}{2}m_V^2 V_0^2+\frac{1}{2}m_b^2 b_0^2+3 \Lambda_V( g_V V_0)^2(g_b b_0)^2.
\label{eq:RMF_EDF}
\end{eqnarray}
where, ($f_{s,t},\ \overline{f}_{s,t}$) are the spin and isospin dependent fermi distribution functions for the nucleon and anti-nucleon respectively. 
And, ($\phi_0,\ b_0,\ V_0$) are the ground state expectation values of the auxiliary fields. The expectation value $\phi_0$
is obtained by solving
\begin{eqnarray}
&&\hspace{-.3in}\left(\frac{m_S}{g_S}\right)^2(g_S \phi_0)+\frac{\kappa}{2}(g_S \phi_0)^2+\frac{\lambda}{6}(g_S \phi_0)^3 \\ \nonumber
&=& \sum_{s,t}\int\frac{d^3k}{(2 \pi)^3}\frac{M_{t}-g_S \phi_0}{\sqrt{k^2+(M_{t}-g_S \phi_0)^2}}(f_{s,t}+\overline{f}_{s,t}), \\ \nonumber
\end{eqnarray}
and the vector ground state expectation values are obtained by solving a pair of coupled equations:
\begin{eqnarray} \nonumber
\rho_p+\rho_n&=&\left(\frac{m_V}{g_V}\right)^2(g_V V_0)+2 \Lambda_V(g_b b_0)^2(g_V V_0) \\ \nonumber
&+&\frac{\zeta}{6}(g_V V_0)^3 \\
\rho_p-\rho_n&=&\left(\frac{m_b}{g_b}\right)^2(g_b b_0)+2 \Lambda_V(g_V V_0)^2(g_b b_0). \\ \nonumber
\end{eqnarray}
For a detailed description of the main features and how they match to supernovae and neutron star observations 
see Ref.\ \cite{Steiner2013}. 
The RMF models we focus on in this paper are the most commonly used in astrophysical calculations: FSUgold \cite{FSUgold,*FSUgold_original}, IUFSU \cite{IUFSU}, TM1 \cite{TM1}, TMA \cite{TMA},
DD2 \cite{DD2}, SFHx \cite{AndrewEOS}, SFHo\cite{AndrewEOS}.
Similar to the work on Skyrme models, in Ref.\ \cite{DutraRMF} a full analysis of 263 different RMF 
parametrizations have been carried out with the same set of $11$ infinite matter constraints,
and only the Z271v5 and Z271v6 \cite{Horowitz2002} parametrizations were found to meet the criteria. Hence, 
we also include them in our analysis.

\subsection{Zero Temperature}
\label{zerot}

\begin{center}
\begin{figure}[t]
\includegraphics[scale=0.35]{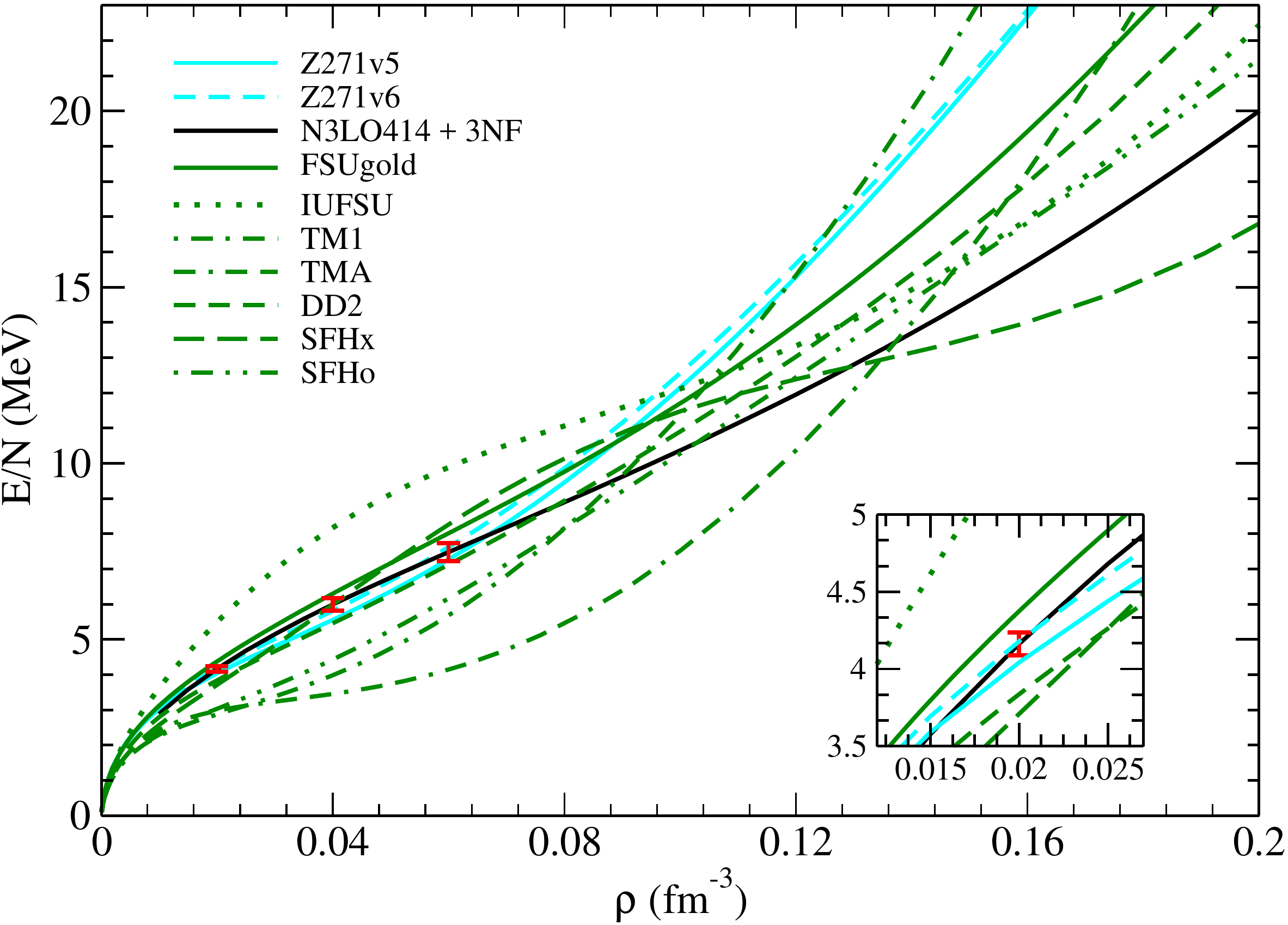}
\caption{(color online) Energy per particle for neutron matter at $T = 0$, where the red bars show the 
constraints derived in the previous sections. In the inset we show a close-up of the low-density
($\rho \simeq 0.02$ fm$^{-3}$) regime. }
\label{fig:RMF_T0Y0}
\end{figure}
\end{center}

\begin{figure}[t]
\includegraphics[scale=0.35]{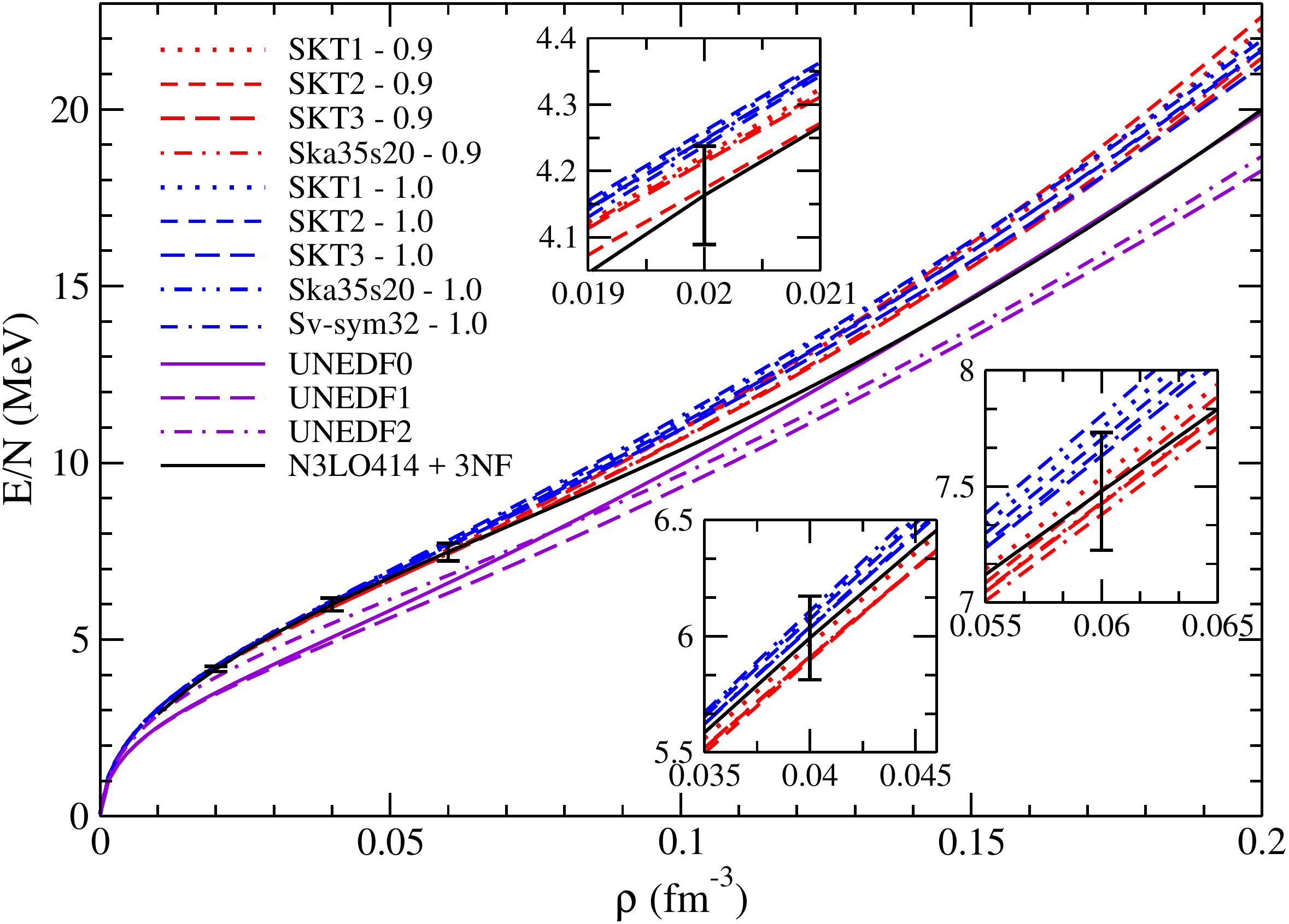}
\caption{(color online) Equation of state for pure neutron matter. The insets show the uncertainty 
bands from neglected correlations in the perturbative treatment from Section \ref{benchmark}.}
\label{fig:FTOY0}
\end{figure}

As the purpose of this work is to gradually select among the many available mean field models those that can be used to 
understand the thermodynamic evolution of a binary merger or core collapse supernova and are consistent with our current knowledge of
baryonic matter, we proceed as follows:

In figures \ref{fig:RMF_T0Y0} and \ref{fig:FTOY0} we select among relativistic and skyrme models respectively based on our findings from 
previous sections by comparing to the chiral \nnnlo~414 interaction with NN and 3NF forces. We have compared our constraints with 
\cite{BrownAchim} and have found that our values are well within the bands in that work.
For the remaining models, we employ constraints from Dutra et al.\ \cite{Dutra2012}, and since 
the refitted Skyrme parametrizations selected in Ref.\ \cite{BrownAchim} are essentially new parametrizations, they need to be tested 
for validity. 

We observe in Fig.\ \ref{fig:RMF_T0Y0} that phenomenological models have wide variations,
even down to very low densities, in contrast to the microscopic predictions shown in
Fig.\ \ref{Fig_nmat_rel}. This model dependence can be traced to the weak correlation between
the low-density neutron matter equation of state and the observables used to constrain
the phenomenological models.
Among the relativistic models only Z271v6 agrees with the 
uncertainty bands from microscopic calculations and therefore
is the only parametrization considered any further in this analysis. 
We notice however that both the DD2 and SFHx parametrizations are very close to fulfilling the 
low-density constraints, and therefore we will test them against 
the set of constraints from Dutra et al.\ \cite{Dutra2012} in the following. 
For the non-relativistic models all the UNEDF parametrizations fail the low density constraints. 
We note however that it should be possible to include low density 
neutron matter calculations as done in Ref.\ \cite{BrownAchim} to achieve a better agreement.
Although the low-density regime is not important for some neutron star properties, for instance 
the mass-radius relation, it is essential for analyzing electron neutrino transport during core-collapse 
supernovae.

 In Fig.\ \ref{fig:FTOY0} we show the chiral equation of state and 
low-density uncertainty estimates together with the various Skyrme models 
that satisfy also the infinite matter constraints.
The uncertainty band for neutron matter at $\rho = 0.02\ \text{fm}^{-3}$ excludes parametrizations
SKT1-1.0, SKT2-1.0, SKT3-1.0, Ska35s20-1.0, Sv-sym32-1.0. However, the Ska35s20-1.0 
Skyrme interaction just misses the low-density constraint, and we choose to consider it in further
analyses as the best representative of the models with $m^*/m = 1.0$. 

In the high-density 
regime the differences between mean field models and the chiral equation of state increase. 
For instance, at saturation density
the variations in Skyrme parametrizations are between $5\%$ and $10\%$ more repulsive than the 
perturbative prediction, while RMF models can differ by up to $40\%$. 
At these densities and beyond, theoretical error bands on the perturbative calculation increase
due to the approximate treatment of three-body forces in the two-body normal-ordering approximation
\cite{bogner05,holt09,holt10,hebeler10} at second-order in perturbation theory, whose uncertainties have
not been evaluated in the present work (but see Ref.\ \cite{kaiser12} for an initial study).
We therefore do not place strong constraints on the mean field models from chiral effective field theory 
beyond low densities. We note however that given the results obtained at low densities a 
$5\%$ error is to be expected and we can therefore assume that the parametrizations SKT1-0.9, SKT2-0.9,
SKT3-0.9, Ska35s20-0.9 and Ska35s20-1.0 are in very good agreement with the results obtained from 
chiral interactions.

\begin{figure}[t]
\includegraphics[scale=0.35]{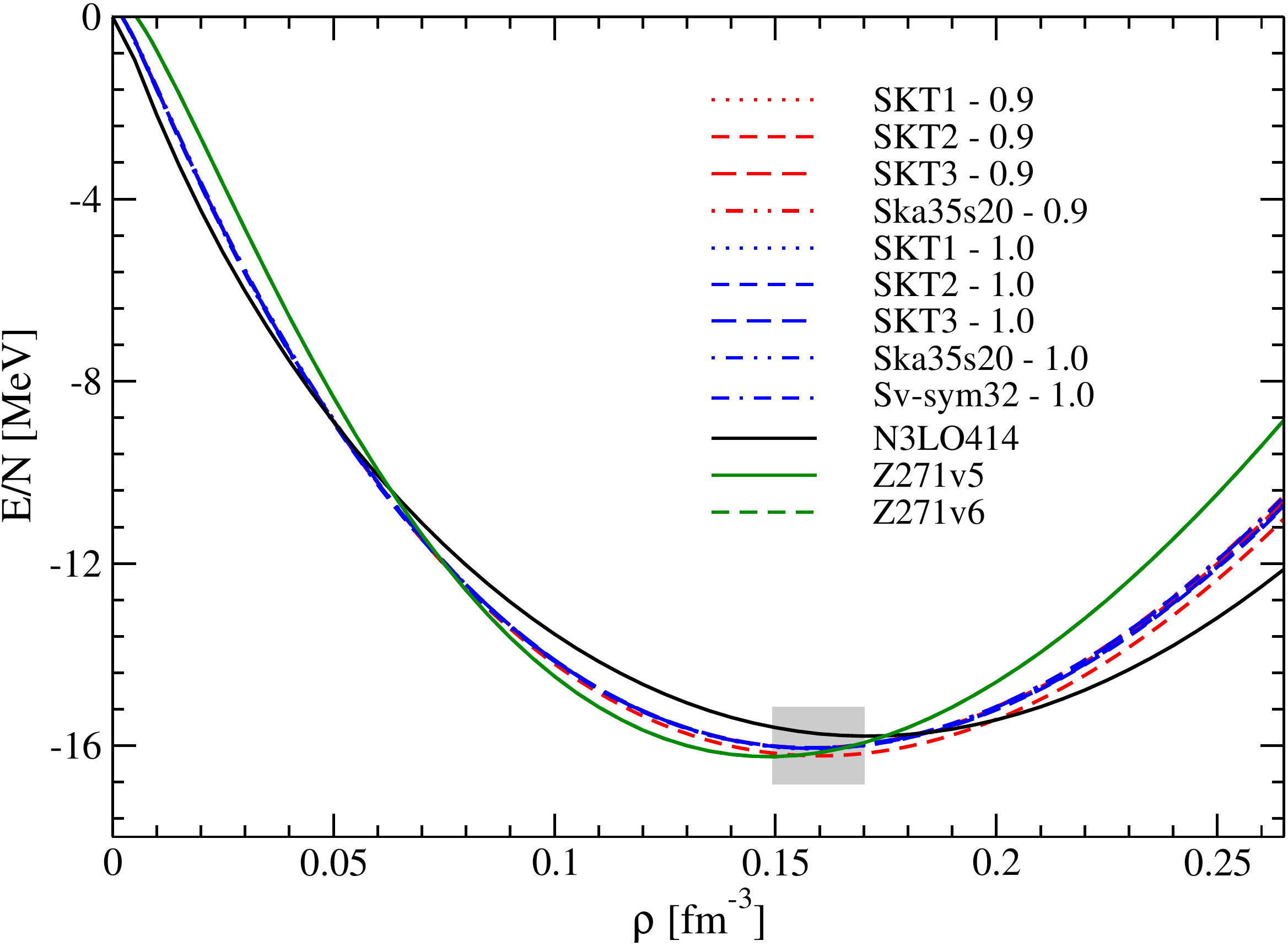}
\caption{(color online) Same as Fig.\ \ref{fig:FTOY0} but for symmetric nuclear matter. We 
allow a 5\% tolerance at nuclear matter saturation density shown by the grey box.}
\label{fig:FTOY05}
\end{figure}
\begin{table*}[t]
\begin{center}
 \begin{tabular}{|c|c|c|c|c|c|c|c|c|}
  \hline
 \bf Name & compatible &$K_m \ (\text{MeV})$  & $K'\ (\text{MeV}) $ & $J\ (\text{MeV})$ & $L\ (\text{MeV})$ & $K_{\tau,v}\ (\text{MeV}) $& $S(\rho/2)/J$& $3P_{PNM}/L \rho_0$ \\
  \hline
  \hline
  Ska25s20 - 0.9 & - &220& 413& 32.1& 50.9& -345& 0.66& 1.0 \\
  \hline
  Ska35s20 - 0.9 & + & 240& 378& 32.2& 53.6& -374& 0.65& 1.0 \\
  \hline
  SKT1 - 0.9    & + &238& 382& 32.6& 55.3& -378& 0.65& 1.0\\
  \hline
  SKT2 - 0.9    & + &240& 385& 33.1& 58.0& -385& 0.64& 1.0\\
  \hline
  SKT3 - 0.9    & + &237& 382& 32.0& 52.6& -370& 0.65& 1.0\\
  \hline
  Sv-sym32 - 0.9 & - &234& 384& 31.5& 49.5& -360& 0.66& 1.0 \\
  \hline
  \hline
  Ska25s20 - 1.0 & - & 220& 415& 32.0& 46.4& -355& 0.67& 1.0 \\
  \hline
  Ska35s20 - 1.0 & + &240& 379& 32.2& 50.6& -385& 0.66& 1.0 \\
  \hline
  SKT1 - 1.0     & + & 237& 384& 32.5& 51.5& -386& 0.66& 1.0\\
  \hline
  SKT2 - 1.0     & + & 236& 387& 32.3& 50.4& -381& 0.66& 1.0\\
  \hline
  SKT3 - 1.0     & + & 237& 385& 33.0& 48.9& -377& 0.66& 1.0\\
  \hline
  Sv-sym32 - 1.0 & + & 237& 375& 32.0& 47.7& -387& 0.66& 1.1\\
  \hline
  \hline
  DD2 & - & 243& -169& 31.7& 55.0& -461.7& 0.66 & 1.0 \\
  \hline
  SFHx & - & 239& 457& 28.7 & 23.2& -- & -- & 1.0\\
  \hline
  Z271v5 & + & 270& 734& 34.0& 73.9& -389& 0.57& 1.0\\
  \hline
  Z271v6 & + & 270& 734& 33.8& 70.9& -388& 0.57& 1.0\\
  \hline
  \hline
  N3LO414 + 3NF  & + & 223& 270& 32.5& 53.8&-424 & 0.70& 1.0\\
  \hline
  \hline
  \bf Range of constraint &  & 190-270 & 200-1200 & 30-35 & 40-76 & -760 to -372&0.57-0.86 & 0.9-1.1\\
  \hline
  \hline
\end{tabular}
\caption{Constraints on the properties of infinite nuclear matter from Dutra et al.\ \cite{Dutra2012,DutraRMF}. 
See text for definitions and details.}
\label{table:constraints}
\end{center}
\end{table*}

For the symmetric nuclear matter equations of state shown in Fig.\ \ref{fig:FTOY05} 
the predictions of different models are very similar, reflecting the fact that the 
parametrizations have been predominantly fitted to properties of infinite matter around and below the 
saturation density. 
As discussed in Section~\ref{benchmark}, in this case we do not have a tight constraint on the
perturbation theory uncertainties, and considering an average error of $\Delta E / E \approx 10\%$,
all parametrizations in Fig.~\ref{fig:FTOY05} are compatible with the perturbative results. 
As a further check we compare the predictions for the saturation point and find that all parametrizations 
are able to reproduce the empirical result within the $5\%$ box shown in Fig.~\ref{fig:FTOY05}. We therefore 
cannot use the $T=0$ SNM equation of state to further eliminate mean field models.

In Table \ref{table:constraints} we summarize the numerical infinite matter constraints from 
Ref.\ \cite{Dutra2012}. In particular, the 
observables listed are the symmetric matter incompressibility
\be
K_m = 9\left . \frac{\partial P}{\partial \rho}\right |_{\rho_0}
\ee
at saturation density, where $P = \rho^2 \frac{\partial (E/N)}{\partial \rho}$ is the pressure; 
the derivative of the incompressibility with respect to density (the so-called skewness parameter)
\be
K' = -27 \rho_0^3 \left . \frac{\partial^3 (E/N)}{\partial \rho^3}\right |_{\rho_0};
\ee
the symmetry energy
\be
J = S(\rho_0) = \left . \frac{\partial(E/N)}{\partial \delta_{np}^2}\right |_{\rho_0},
\ee
where $\delta_{np} = \frac{\rho_n - \rho_p}{\rho_n + \rho_p}$ is the isospin asymmetry;
the symmetry energy slope
\be
L = 3 \rho_0 \left . \frac{\partial S}{\partial \rho}\right |_{\rho_0};
\ee 
the isospin incompressibility 
\be
K_{\tau,v} = K_{\text{sym}} - L \left ( 6 + \frac{K^\prime}{K_m} \right ) ,
\ee
where $K_{\text{sym}} = 9 \rho_0^2 \left . \frac{\partial^2S}{\partial \rho^2}\right |_{\rho_0}$;
the ratio between the symmetry energy at half saturation density and its value at 
$\rho_0$; and finally the quantity $3 P_{PNM}/L \rho_0$, where $P_{PNM}$ is the pressure 
in PNM at saturation density.
In addition, there are four ``band constraints'' on the symmetric
nuclear matter and pure neutron matter equations of state 
in specific density regimes below and above nuclear saturation:
\begin{enumerate}
 \item $P_{\text{SNM}}$ for $1.2<\frac{\rho}{\rho_0}<2.2$
 \item $P_{\text{SNM}}$ for $2<\frac{\rho}{\rho_0}<4.6$
 \item $P_{\text{PNM}}$ for $2<\frac{\rho}{\rho_0}<4.6$
 \item $(E/N)_{\text{PNM}}$  for $0.014<\frac{\rho}{\rho_0}<0.106$.
\end{enumerate}

First we note that the two RMF parametrizations that were close to reproduce low--density neutron matter constraints,
DD2 and SFHx, do not satusfy the constraints. In particular DD2 has a skewness parameter $K'$ with the wrong sign ($K'=-168.7$\,MeV) while 
SFHx has too low a value for the slope $L$ of the symmetry energy ($L = 23.18$\,MeV).
Furthermore, as can be seen from Table \ref{table:constraints}, some of the new parametrizations
are inconsistent with the empirical constraints in Ref.\ \cite{Dutra2012}. Specifically, Ska25s20-0.9, 
Ska25s20-1.0 and Sv-sym32-0.9 fail to meet the
constraint for isospin incompressibility. 

We note that the discrepancies between our results 
for the symmetry energy and the 
ones reported in Table I from Ref.\ \cite{BrownAchim} are due to different
definitions. In our case we follow Ref.\ \cite{Dutra2012} and start with the 
generic expression at arbitrary isospin asymmetry $\delta_{np} = \frac{\rho_n-\rho_p}{\rho}= 1- 2y_p$ and 
define $S$ as the second derivative of the energy per particle with respect to $y_p$, as shown previously, 
while in Ref.\ \cite{BrownAchim} $S$ is 
obtained as the difference between the PNM and the SNM equations of state at a given density. 
For microscopic approaches (like MBPT) these two definitions of $S$ give very similar values 
(see e.g., \cite{Drischler2014,Gandolfi2014}).

As mentioned above the chiral 414 MeV interaction is used at second order in many-body 
perturbation theory with two- and three-body forces. Results using the potential with a cutoff of 
450 MeV are very close to those from the \nnnlo~414 potential. We see that in all cases the 
microscopic calculations of the equation of state with this potential are consistent with the empirical
infinite matter constraints considered in Ref.\ \cite{Dutra2012}. We note that all models considered 
meet the band constraints.

\subsection{Finite Temperature}
\label{thermodynamics}

In this section we focus on the nuclear thermodynamic equation of state of homogeneous 
matter from the mean field models that were found in the last section to be compatible with the 
zero-temperature equation of state
from low-momentum chiral nuclear forces and analyze the conditions necessary for consistency at finite
temperature. We therefore focus on only the parametrizations SKT1-0.9, SKT2-0.9, SKT3-0.9, Ska35s20-1.0,
Ska35s20-0.9 and Z271v6. Since there are no nonpertubative microscopic calculations available, 
we cannot assign uncertainty bands at specific densities as in the zero-temperature analysis. However, one 
would not expect the many-body perturbation expansion to change dramatically by including finite-temperature
effects, and therefore we use the chiral interaction N3LO414 together with 3NF forces to be a qualitative guiding 
tool for low-density thermodynamical properties. 
\begin{table*}[tbh]
\begin{center}
  \begin{tabular}{| c| c| c| c| c| c| c| c|}
  \hline
 \bf Name & ${\alpha}$ &{\bf a }($\text{MeV}\ \text{fm}^{5}$) &{\bf b} ($\text{MeV}\ \text{fm}^{5}$)& {$\mathbf{t_0}$} ($\text{MeV}\ \text{fm}^{3}$) & $\mathbf{x_0}$& $\mathbf{t_3}$ ($\text{MeV}\ \text{fm}^{3(1+\mathbf{\alpha})}$) &{$\mathbf{x_3}$}  \\
  \hline
  \hline
  Ska35-0.9 &0.35 &-172.485& 172.087& -1767.71& 0.282732& 12899.2 & 0.413266\\
  \hline
  SKT1-0.9 &1/3 & -112.324& 142.467& -1810.72& 0.28223& 12863.0& 0.392585\\
  \hline
  SKT2-0.9 &1/3 &-113.857& 143.999& -1807.87& 0.267778& 12802.4& 0.366144 \\
  \hline
  SKT3-0.9 &1/3 & -124.432& 148.492& -1812.16& 0.288584& 12906.6& 0.416129\\
  \hline
  Ska35-1.0 &0.35 &-2.41114& -0.507978& -1767.92& 0.247025& 12910.2& 0.220377\\
  \hline
   \hline
\end{tabular}
\caption{Parameters for the refitted Skyrme interactions of Ref.\ \cite{BrownAchim}.}
\label{tab:cf}
\end{center}
\end{table*}
The aim is to identify mean field models that quantitatively reproduce 
chiral nuclear thermodynamics at low to moderate densities and temperatures. 
The coefficients for the refitted Skyrme parametrizations that meet our selection criteria so far 
are given in Table \ref{tab:cf}.


The mean field models we are considering are fitted to nuclei and infinite matter at $T=0$ and 
contain no empirical input at finite temperature. The nucleon effective mass however controls the density
of states and therefore aspects of thermal excitations. We anticipate that the single-particle 
properties of nucleons in mean field models will control whether the associated thermodynamic
equations of state will be consistent with those from perturbative chiral interactions. 
We choose low- and high-temperature representative values $T=5,25$\,MeV. 

The formalism employed in this section is self-consistent Hartree-Fock theory which treats the density 
matrix as effectively of one-body type. The problem of treating interactions properly is cast into an 
energy density functional for which a corresponding local density is found by minimization of the energy. 
The Skyrme and RMF models fall into this category. 
In the thermodynamic limit an ensemble equivalence is guaranteed, and the canonical or grand-canonical 
potential can be minimized.
As we opt to have temperature $T$ and density $\rho$ as external fixed parameters, 
we work in the canonical ensemble. The set of self-consistent equations to be satisfied at Hartree-Fock level is
given by Eqs.\ (\ref{eq:sceq1}) and (\ref{eq:sceq2}) below.
For a given momentum-dependent nucleon single-particle spectrum $\epsilon_{s,t}(k)$,
where $s$ and $t$ are the spin and isospin quantum numbers, the energy density and entropy density
can be calculated from
\begin{eqnarray}
\label{eq:sceq1}
f_{s,t}(k)&=&\bigg [1+ e^{\big(\epsilon_{s,t}(k)-\mu \big)/T}\bigg ]^{-1},\\ \nonumber
\rho &=& \sum_{s,t}\int \frac{d^3k}{(2\pi)^3}f_{s,t}(k),\\ \nonumber
\tau &=& \sum_{s,t}\int \frac{d^3k}{(2\pi)^3}k^2f_{s,t}(k),\\ \nonumber
&&\hspace{-.57in} S/V = -\sum_{s,t} \int \frac{d^3k}{(2\pi)^3} \big[f_{s,t} \ln f _{s,t}+ (1-f_{s,t})\ln (1-f_{s,t})\big],\\ \nonumber
\label{eq:entropy}
\end{eqnarray}
where a sum over all discrete quantum numbers is performed (spin and isospin). 
The chemical potential can be found by inverting the expression for the density.
From the energy density functional, an effective mass and mean field shift can be extracted:
\begin{equation}
\begin{split}
M^{*}_t&=\frac{1}{2}\left (\frac{\delta \mathcal{E}}{\delta \tau_t}\right )^{-1}\\
U_t &= \frac{\delta \mathcal{E}}{\delta \rho_t},
\end{split}
\label{eq:sceq2}
\end{equation}
which gives a single-particle spectrum of the form
\begin{equation}
\epsilon_t(k) = k^2/2M^*_t + U_t.
\end{equation}

\begin{figure}[t]
\includegraphics[scale=0.34]{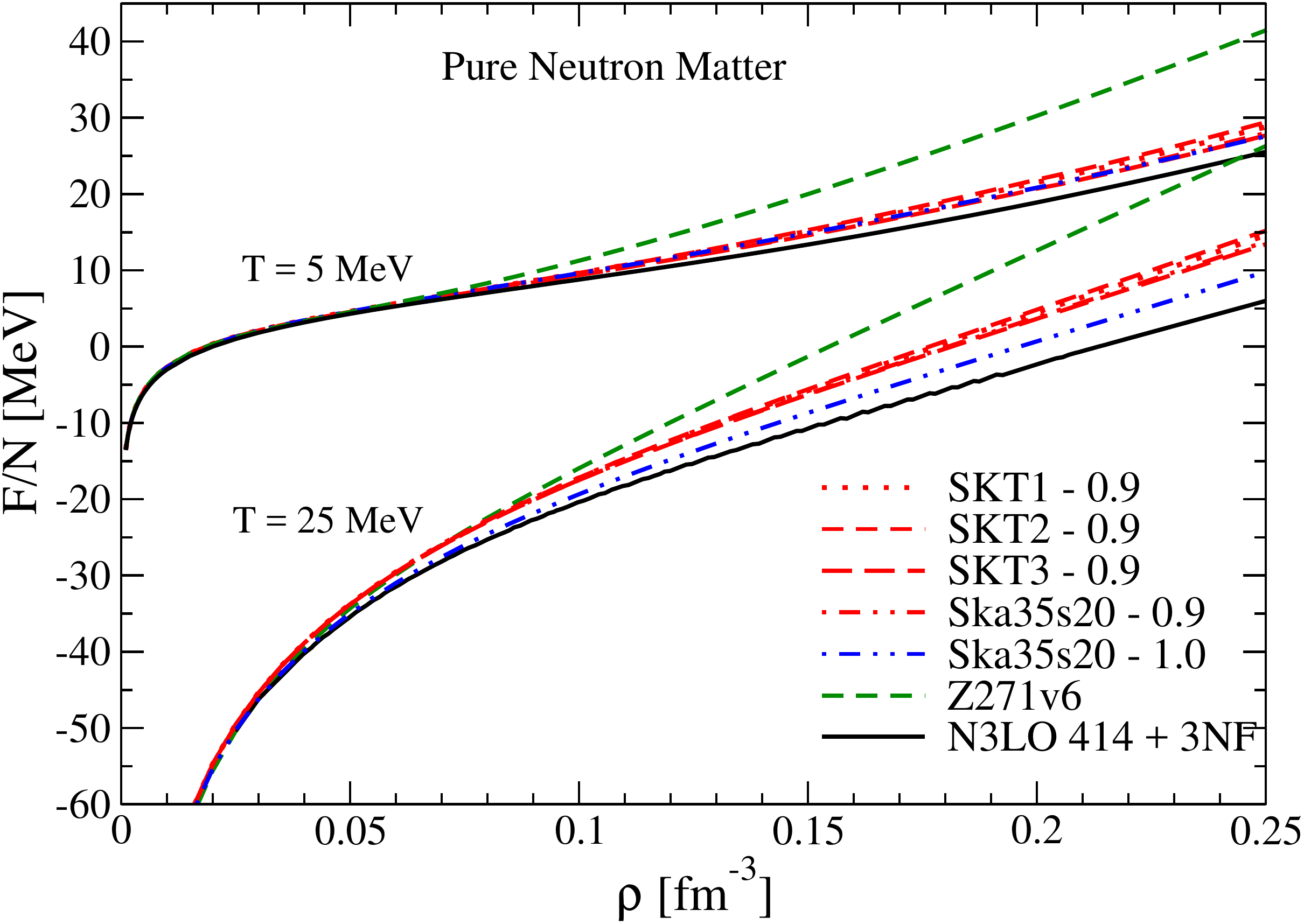}
\caption{(color online) Free energy per nucleon in pure neutron matter at temperatures $T=5,25$\,MeV
for the 414 MeV chiral nuclear potential and mean field models.}
\label{fig:FT5a}
\end{figure}

\begin{figure}[t]
\includegraphics[scale=0.34]{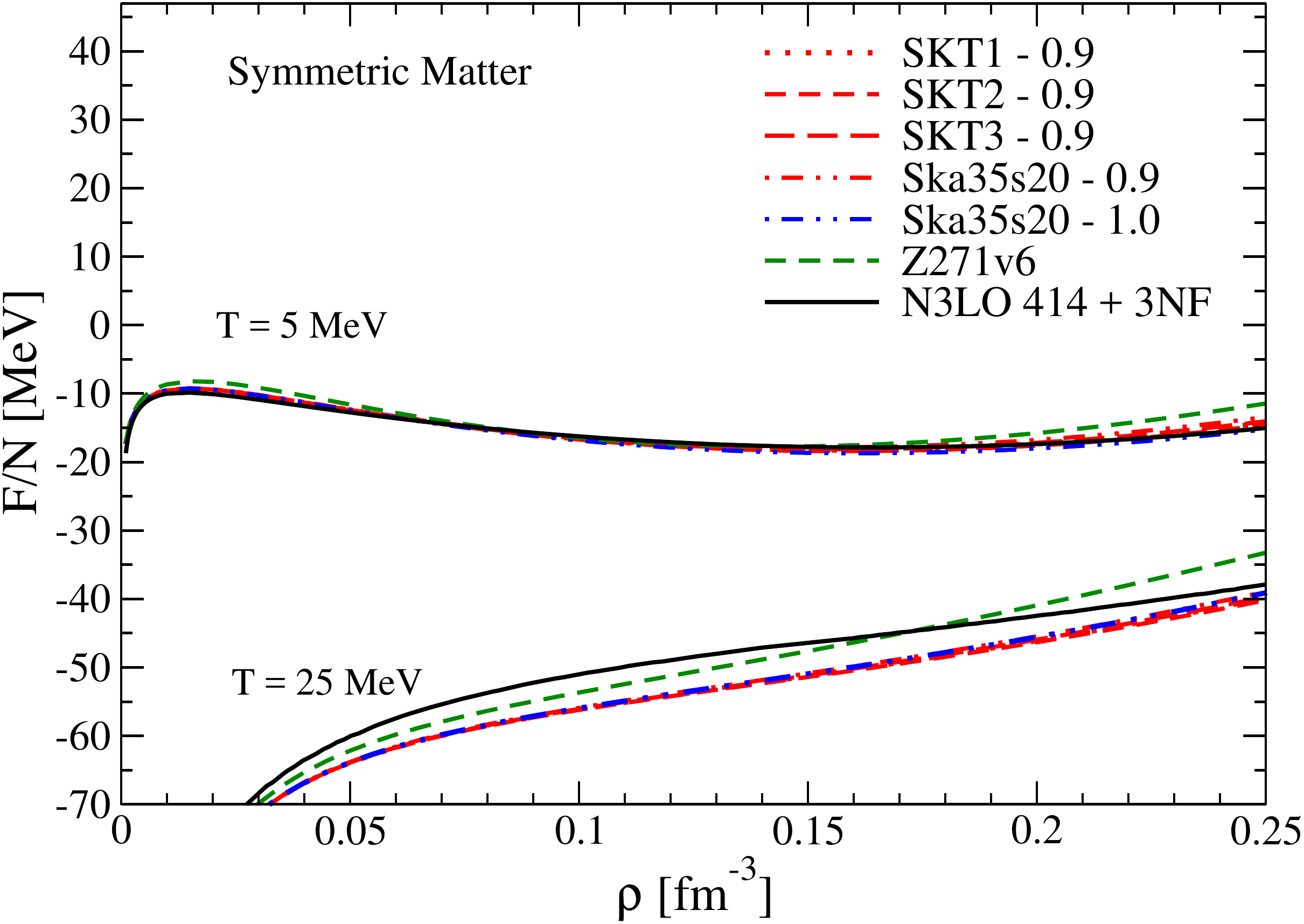}
\caption{(color online) Same as Fig.\ \ref{fig:FT5a} except for symmetric nuclear matter.}
\label{fig:FT5b}
\end{figure}

In Figs.\ \ref{fig:FT5a} and \ref{fig:FT5b} we plot the free energy per particle of neutron matter
and isospin-symmetric nuclear matter for the microscopic and mean field models under consideration.
We employ the Matsubara finite-temperature formalism up to second order in perturbation theory 
to compute the free energy per particle from the chiral two- and three-body potentials. Additional 
details can be found in Refs.\ \cite{Wellenhofer2014,Wellenhofer2015}.
At $T=5$\,MeV the PNM free energy is similar to that at zero temperature, where the mean field equations of
state are consistently less attractive than that from chiral low-momentum interactions. However, 
at $T=25$\,MeV the Skyrme mean field models exhibit larger variations in the free energy. 
In particular smaller values of the effective mass result in a larger kinetic energy contribution and a 
small entropy term, both of which reduce the attraction at high temperatures.
In symmetric nuclear matter at low temperatures there is consistency between all mean field models
considered and the chiral equation of state. At the largest temperature considered ($T = 25$\,MeV) 
the Skyrme models remain in good agreement with each other, owing to the very similar values of the
effective masses, but deviate by about 5 MeV from the chiral effective field theory prediction. 
This pattern is different than that in the neutron matter equation of state, and it suggests that the 
average effective nucleon mass in symmetric nuclear matter is smaller than that in the Skyrme mean field
models.

\begin{figure}[t]
\includegraphics[scale=0.34]{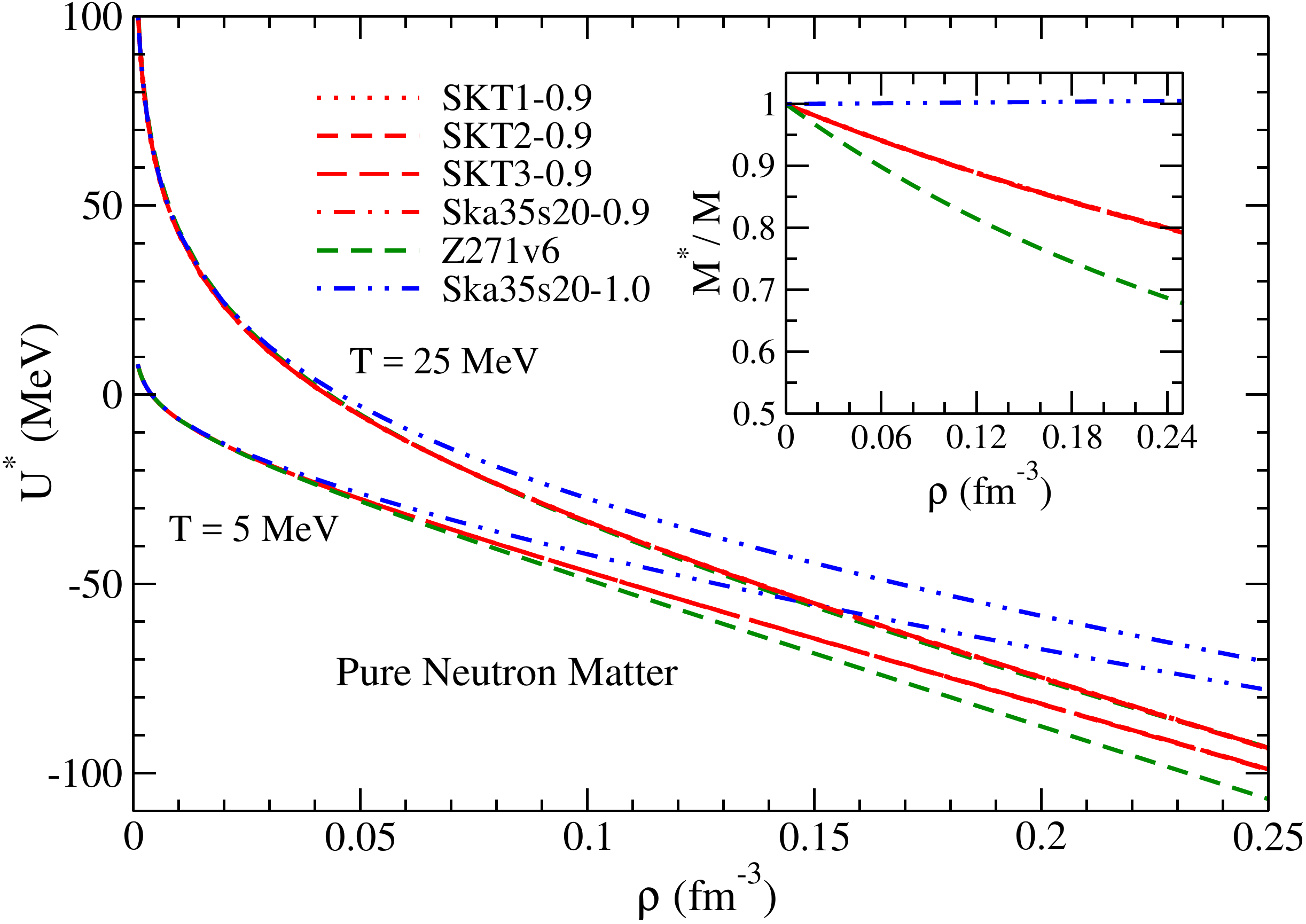}
\caption{(color online) Neutron mean field shift and effective mass in pure neutron matter. Note that all
``Sk...0.9" Skyrme interactions have nearly identical effective masses.}
\label{fig:UandMa}
\end{figure}

\begin{figure}[t]
\includegraphics[scale=0.34]{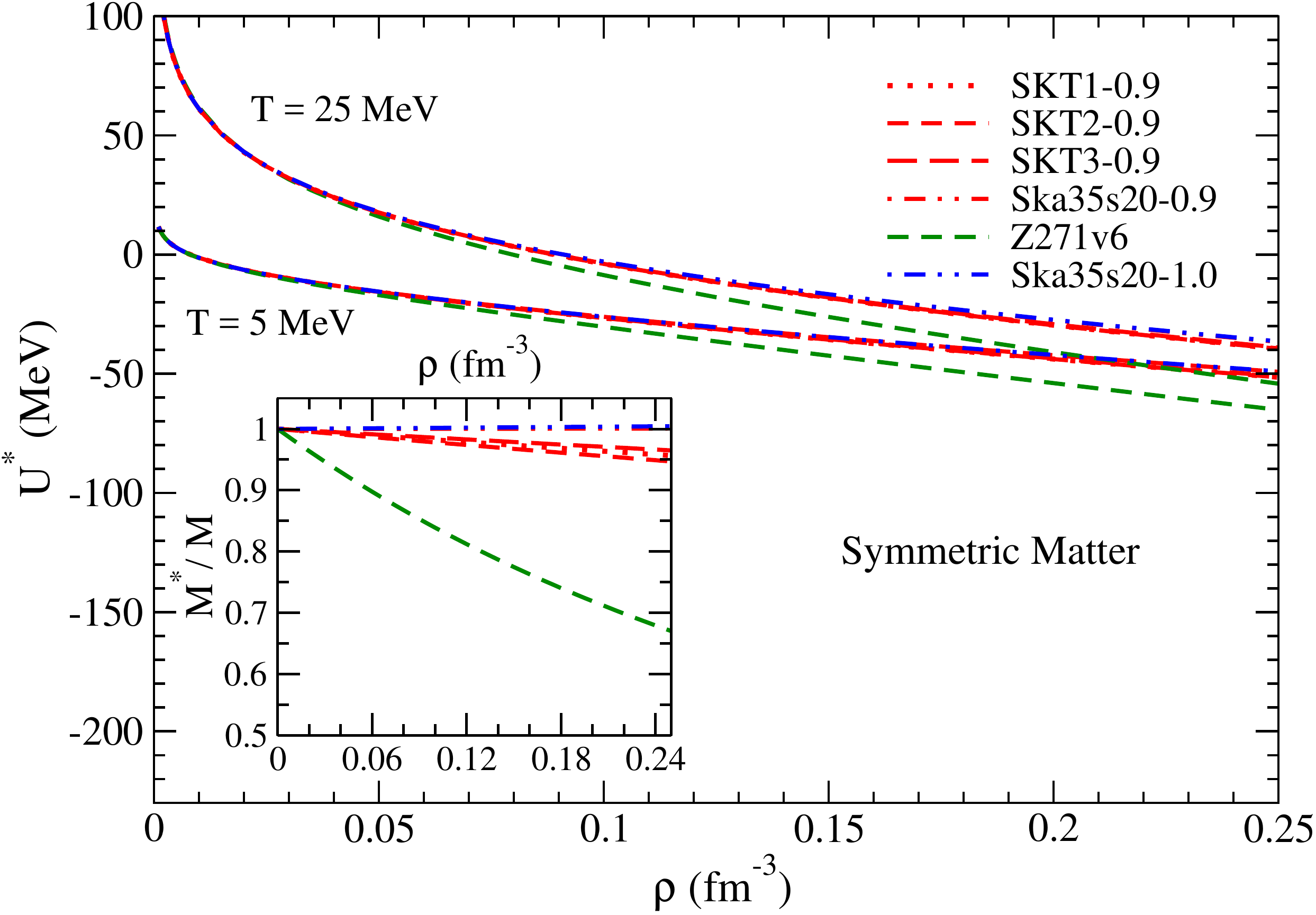}
\caption{(color online) Same as in Fig.\ \ref{fig:UandMa} except for symmetric nuclear matter.}
\label{fig:UandMb}
\end{figure}

To explain the difference more carefully, we study the single-particle spectrum. 
The relativistic and nonrelativistic models treat this quantity (as well as the chemical potential) 
differently. We first compare the non-relativistic reduction 
of the single-particle spectrum in RMF models to that in the Skyrme models:
\begin{equation}
\begin{split}
{e_t}_{\text{Skyrme}}(k) =& \epsilon_t(k)-\mu_t = \frac{k^2}{2M^{*}_t}+U_t-\mu_t^{\text{NR}}\\
\equiv& \frac{k^2}{2 M^{*}_t} + U^{*}_t, \\
{e_t}_{\text{RMF}}(k) =& \epsilon_t(k)-\mu_t = \sqrt{k^2+{M^{*}_t}^2} + U_t -\mu_t^{\text{R}} \\
\approx& \frac{k^2}{2M^{*}_t} + M^{*}_t+U-\mu_t^{\text{R}} \\
\equiv& \frac{k^2}{2M^{*}_t}+U^{*}_t.
\end{split}
\label{eq:sp}
\end{equation}
\hspace{-.052in}At the HF level the low-energy spectrum can be accurately modeled by an effective mass and 
mean field shift, both density dependent. We show in Figs.\ \ref{fig:UandMa} and \ref{fig:UandMb} 
the effective mass and mean field shifts in neutron matter and symmetric nuclear matter 
for the models under consideration.
In the case of SNM, the lower value of the effective mass corresponds to a higher value of the 
energy per baryon for Skyrme interactions, which results from a larger kinetic energy contribution
to the free energy. 
Since the mean field shifts for the different Skyrme interactions with $M^*/M = 0.9$ 
in PNM at $\rho = 0.1$\,fm$^{-3}$ are nearly identical, the kinetic contribution dominates the difference
while the interaction contribution depends mostly on the density. 

In contrast, RMF models predict smaller effective masses and larger (negative) effective mean field shifts. 
In the case of Z271v6
these effects balance in SNM, with the effective mass becoming more important only at higher densities 
(higher values of $F/N$).
However, for PNM, Z271v6 has a very small effective mass which leads the kinetic energy to dominate 
over interactions as 
can be seen by higher values of $F/N$ for the whole density range depicted in fig.~\ref{fig:FT5a}. 
Despite the effective mass and energy shifts being very different between these two models, 
the free energies predicted are relatively similar. To understand the physical relevance of these two parameters,
we have to study physical `observables' which are specifically susceptible to them.
In the next section we focus on neutrino response and thermodynamic evolution, both sensitive to the single-particle spectrum,
to differentiate Skyrme and RMF models.


\subsection{Effect of Single-Particle Properties}
\label{differences}
 
Nuclear thermodynamics is governed by the free energy per nucleon as a function of the density and
temperature. Related quantities such as pressure, entropy and chemical potential are given
in terms of the free energy through standard thermodynamic relations. Dynamical phenomena in
nuclear matter at finite temperature
are strongly related to the single-particle properties of nucleons, and
as shown in the last section the effective mass and mean field shift can be quite different between
two models that nevertheless have similar equations of state. 
Thus, to explore a different set of consistency requirements between mean field models and microscopic nuclear dynamics we employ
an analysis of the neutrino response and isentropic curves for core-collapse supernovae.
We focus on Ska35s20-0.9, Ska35s20-1.0, and Z271v6, which are the ``best fit'' 
parametrizations from nonrelativistic and relativistic mean field models to chiral equations of state 
at finite temperature.

\subsubsection{Neutrino Response}
Charged-current weak reactions are primarily responsible for setting the electron neutrino opacity 
in the warm ($T=5-8$\,MeV) and dilute ($\rho = 0.001-0.01\rho_0$) region of last scattering 
(the neutrinosphere) for neutrinos diffusing from the core of a newly born proto-neutron star
\cite{pinedo12,roberts12,roberts}. 
Owing to the nondegenerate conditions, the fermi distribution functions for neutrons and protons 
are strongly smeared in the vicinity of the chemical potential. The proton and neutron energy 
shifts are primarily responsible for modifications to neutrino and anti-neutrino mean free paths,
but the momentum dependence of the single-particle potential can also be relevant for low-density
moderate-temperature conditions.
The inverse mean free path of $\nu_e$ from the reaction $ \nu_e + n
\rightarrow e^{-} + p$ follows from Fermi's golden rule and is given
by
\begin{eqnarray}
 \lambda^{-1}_{\nu_e} &= &\frac{2}{(2\pi)^5}\int d^3p_n\, d^3p_e\, d^3p_p~ \mathcal{W}_{fi}\\ \nonumber
 & \times & \ {\delta}^{(4)}(p_{{\nu}_e} +p_n - p_e - p_p) f_n (1-f_e) (1-f_p),
\label{dsigma}
\end{eqnarray}
where $f$ is the fermi distribution function and ${W}_{fi}$
is the transition probability. Both depend on the single-particle spectrum, and due to the 
energy-momentum conserving delta function, the phase space is also highly impacted by $M^{*}$ and 
$U^{*}$. 

We consider $\beta$-equilibrated matter at the density $\rho = 0.02$\,fm$^{-3}$
and temperature $T=8$\,MeV. We show in Table \ref{umstar} the proton and neutron effective masses and energy 
shifts for the mean field models considered as well as for the chiral n3lo414 potential at Hartree-Fock level 
and the pseudo-potential which resums iterated ladder diagrams (without Pauli blocking) to all orders
in perturbation theory and which is defined in terms of nucleon-nucleon scattering phase shifts (see
Ref.\ \cite{Rrapaj2015} for additional details). The effective mass and mean field shift for the chiral interactions 
come from a global fit of the single-particle spectrum from $k = 0$ to $k=2k_f$.
Note that for the temperature and density considered here, the two Skryme
parametrizations Ska35s20-0.9 and Ska35s20-1.0 give nearly identical single-particle properties.

\begin{figure}[t]
\includegraphics[scale=0.3]{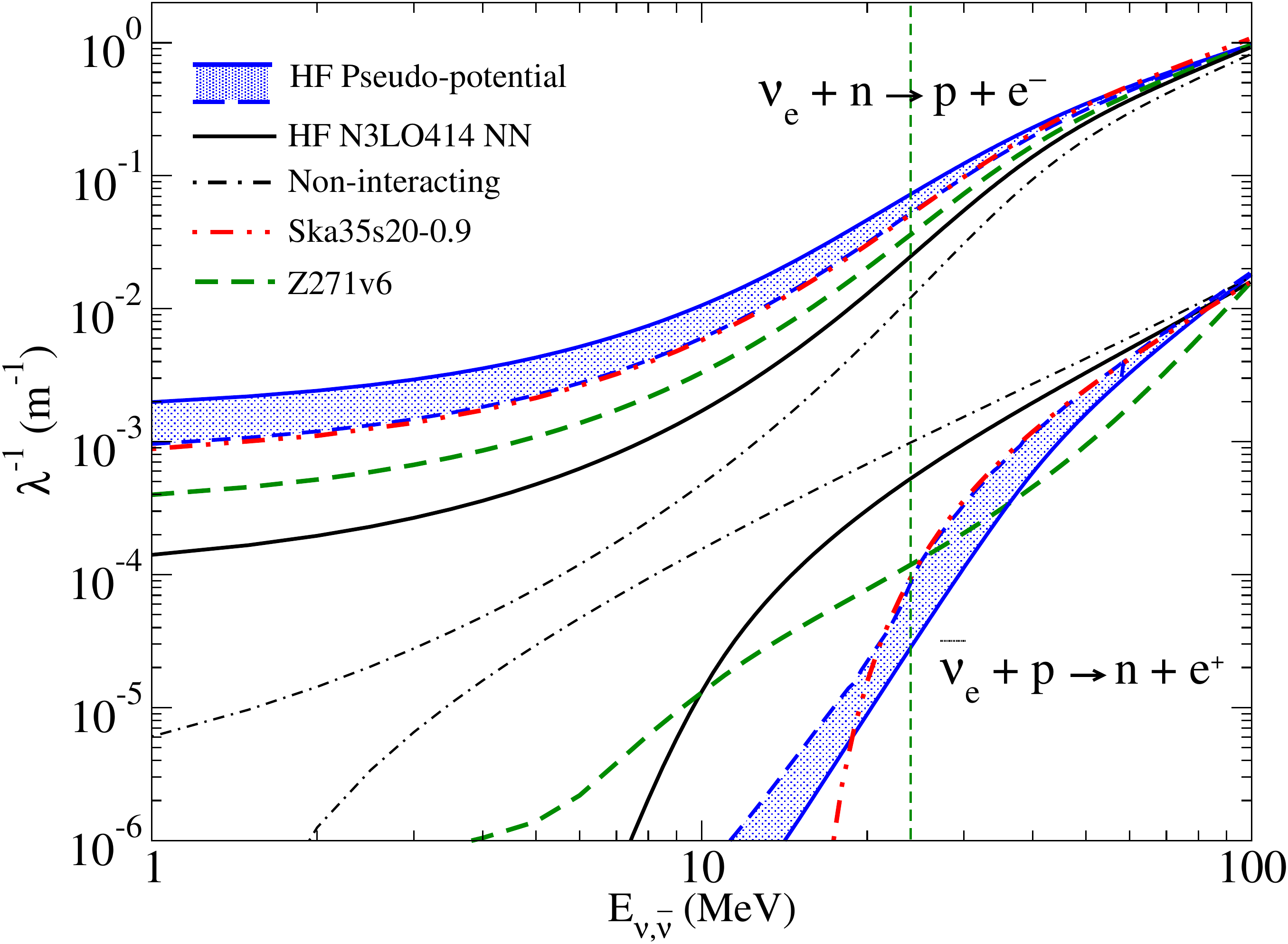}
\caption{(color online) Charged-current rates at $\beta$ equilibrium: $T = 8\ \text{MeV}$, $\rho = 0.02\ \text{fm}^{-3}$. 
The vertical dashed line indicates the `thermal' energy, $E_{\nu_e, \overline{\nu}_e} = 3\ T$.}
\label{fig:lambda}
\end{figure}

Given the paramount importance of charged-current absorption rates
in determining the decoupling regions, and thus the energy spectra of $\nu_e$ and $ \overline{\nu}_e$, 
we plot in Fig.\ \ref{fig:lambda} the opacity from two mean 
field equations of state. 
\begin{table}[h]
\centering
\resizebox{\columnwidth}{0.05\textheight}{%
\begin{tabular}{@{} |c|c|c|c|c|c|c| @{}} 
\hline 
Model & $Y_p $ & $M^*_n/M_n\ $ & $M^*_p/M_p\ $ & $-U_n\ $ & $-U_p\ $ & $\Delta U\ $ \\
\hline 
HF Pseudo-potential & 4.9\% & 0.65 & 0.42 & 22 & 55 & 33\\
HF Pseudo-potential (mod)\,\, & 3.8\% & 0.78 & 0.57 & 18 & 42 & 23\\
HF N3LO414 & 2.2\%  & 0.95 & 0.89 & 8 & 16 & 8  \\
RMF: Z271v6 & 2.8\% & 0.96 & 0.96 & 9 & 22 & 13 \\
Skyrme: Ska35s20-0.9 & 3.3\% & 0.98  & 1.0 & 9& 26 & 17\\
\hline
\end{tabular}
}
\caption{The Hartree-Fock (HF) effective masses $M^*$ and energy shifts $U$ (in units of MeV) for 
protons and neutrons in beta equilibrium
at $n_B = 0.02$\,fm$^{-3}$ and temperature $T = 8$\,MeV.
The difference in proton and neutron mean-field shifts is given by 
$\Delta U = U_n - U_p$, and the proton fraction is denoted by $Y_p$.}
\label{umstar}
\end{table}
We observe that for electron neutrino absorption all mean field models 
are consistent with the lower bound provided by the chiral Hartree-Fock
approximation and the upper bound set by the pseudo-potential. The Skyrme Ska35s20-0.9 
mean field prediction is in good agreement with the pseudo-potential results, given that both models
predict large values of $\Delta U = U_n - U_p$. Since the chiral EFT calculation is only at the self-consistent
Hartree-Fock level, considerable attraction in the iterated tensor force channel is missing, and 
results much closer to those from the pseudo-potential are expected when second-order 
perturbative contributions are included. We plan to 
study these effects more carefully in future work.
To completely determine the neutrino opacity, neutral current reactions are needed. 
Particularly for the case of $\overline{\nu}_e$ in the lower regions of the power spectrum, brehmsstrahlung 
absorption rates dominate \cite{pinedo14,bartl14,Rrapaj2015}. 

\subsubsection{Supernova Isentropic Curves}
During the core collapse of a massive star, the entropy per baryon changes from about 1 $k_B$ to 2-3 $k_B$,
and the adiabatic (isentropic) approximation can be used to describe the 
thermodynamic evolution during such a short time scale.
The temperature versus density at constant entropy then provides a prediction of the temperature 
of the core in the initial and final stages.
Since neutrinos are primarily emitted from the high-density region by neutral- and charged-current reactions, 
the temperature of the core plays 
an important role in determining their spectra \cite{connor2013}. At low temperatures both the density of states 
and entropy are expected to be proportional to the nucleon effective mass $M^*$
\begin{eqnarray} \nonumber
N(0) &=& \sum_t\frac{M^*_t{k_f}_t}{\pi^2},\\
S/V &=& \sum_t\frac{M^*_t{k_f}_t}{3}T.
\end{eqnarray}
For the same entropy per baryon, a smaller effective mass at a given density therefore 
translates to a higher temperature.

\begin{figure}[t]
\includegraphics[scale=0.36]{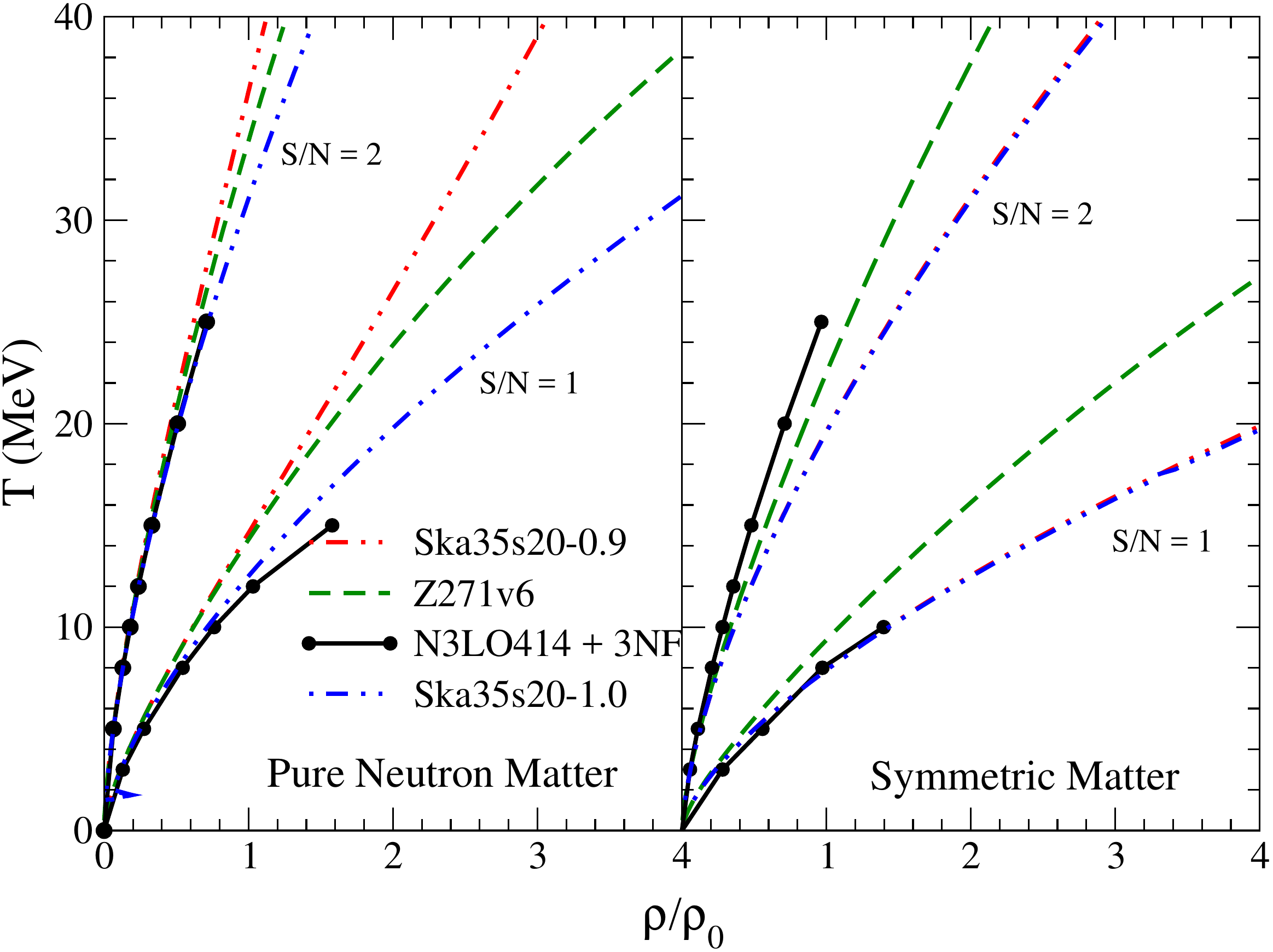}
\caption{(color online) Temperature versus density isentropic lines for PNM and SNM.}
\label{fig:Trho}
\end{figure}

In Fig.~\ref{fig:Trho} we depict the Skyrme, RMF and chiral 
isentropic lines for pure neutron matter and isospin-symmetric 
matter for densities up to 4$\rho_0$.
We use the exact expression for entropy from Eq.~\ref{eq:entropy} in this plot.
For low-density neutron matter ($\rho<\rho_0/2$), 
the Ska35s20-1.0 Skyrme interaction gives results that are consistent with the n3lo414 
chiral potential, which from Fermi liquid theory \cite{davesneholt15} predicts an effective mass 
$M^*/M \sim 1$. Beyond nuclear matter saturation density, the differences between the 
temperature of neutron matter at $S/N=1$ can vary by up to 8\,MeV for the chiral and
mean field theory equations of state, while for $S/N=2$ all models are nearly consistent.
In the case of symmetric nuclear matter the two Skyrme parametrizations considered have 
very similar effective masses (see Fig.\ \ref{fig:UandMb}) and therefore the isentropic curves are 
nearly identical. At the highest temperatures ($T\gtrsim 20$\,MeV) there are large variations
in the predicted densities. In the microscopic calculation with the chiral n3lo414 potential, one 
may interpret this as a significant reduction of the effective mass 
at finite temperature due to a damping of collective modes \cite{donati94}, which normally
lead to an enhanced effective mass. 
In general for temperatures less than $T=10$\,MeV the chiral isentropic curves are in
good agreement with those from the Ska35s20-1.0 Skyrme interaction.

\section{Neutron Star Mass-Radius Relationship}
An important feature of the `strong' physics of neutron stars is the interdependence of mass and radius,
which is uniquely determined by the EoS and the self-gravity of these compact objects. A 
relatively stiff EoS at high densities is required to generate a maximal mass of 2$M_\odot$, which is currently the
highest mass of an observed neutron star \cite{Antoniadis2013}. 
However, chiral interactions have a momentum cutoff that is comparable to the fermi momentum at
about 2$\rho_0$ in neutron matter, and central densities in neutron stars can reach values of 
5$\rho_0$ or higher. The mean field models considered in the present work can therefore be 
used to probe the very high density regime inaccessible to chiral effective field theory with
coarse-resolution potentials. In our analysis we choose to use the chiral interaction up to the limit 
of its validity without model-dependent extensions.
RMF models by construction remain causal, but Skyrme models can become superluminal at 
very large densities. In the present analysis, however, we find no evidence for this behavior.

\begin{figure}[t]
\includegraphics[scale=0.36]{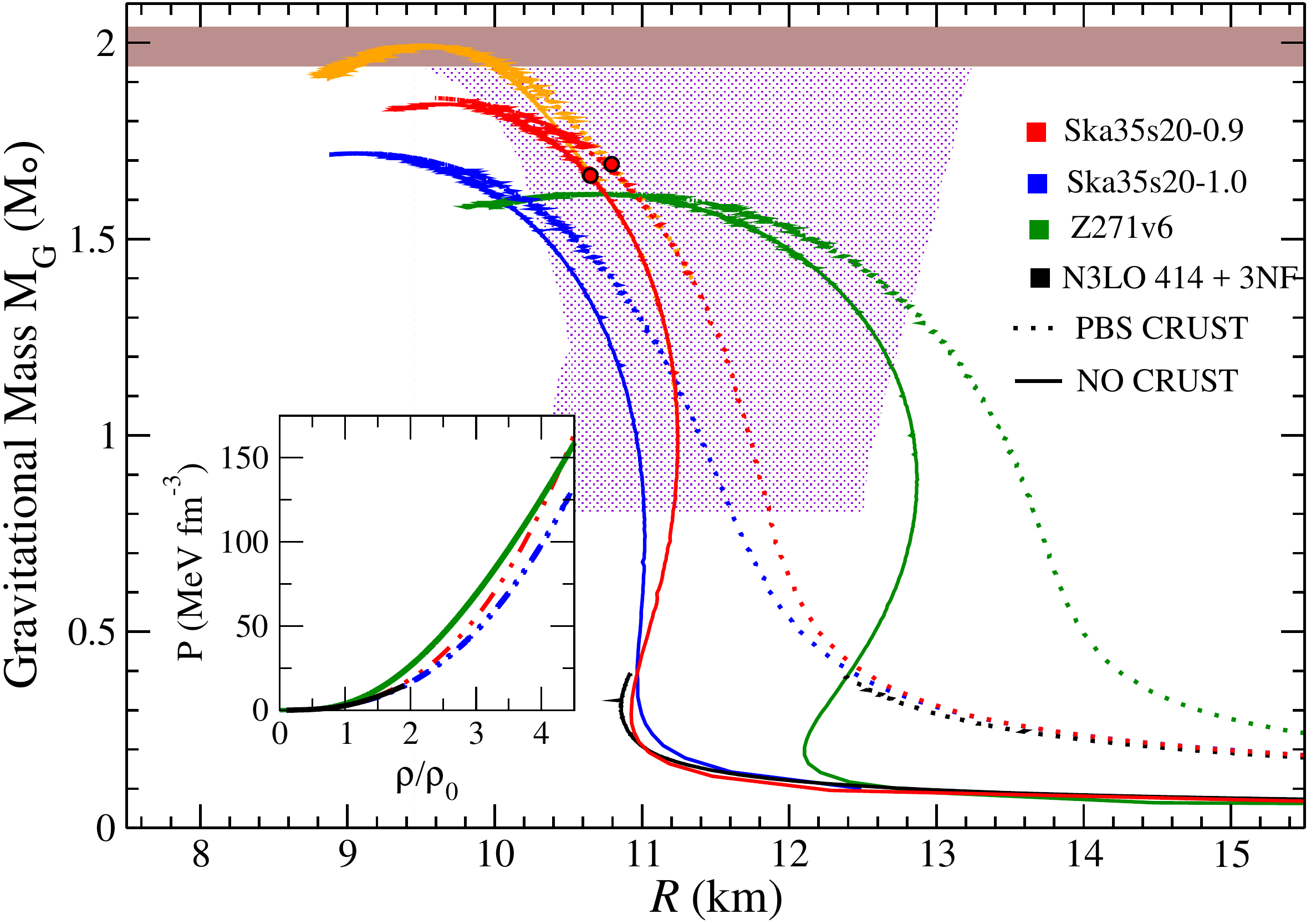}
\caption{(color online) Neutron star mass as a function of radius. The mass-radius observational constraint comes from 
Ref.\ \cite{Steiner2012} and the maximum neutron star mass comes from Refs.\ 
\cite{Demorest2010,Antoniadis2013}. We use the BPS crust EoS \cite{PBS} at low densities. The solid dots indicate the transition region to the constant speed of sound equation of state.}
\label{fig:m-r}
\end{figure}

In Fig.\ \ref{fig:m-r} we plot the neutron star mass vs.\ radius in the absence of protons and 
light leptons at $T = 0$ from the mean field models consistent with the chiral n3lo414 equation of 
state at low to moderate densities. Recently, in Ref.\ \cite{dutra15} a similar analysis has been 
performed for relativistic mean field models consistent with the infinite matter constraints considered
in Section \ref{MFEOS}. Given the low values of the proton fraction found for 
$\beta$-equilibrated 
matter, this approximation is expected to be very good. We show also in Fig.\ \ref{fig:m-r} the current 
observational constraints on the mass and radius \cite{Steiner2012}. For each mean field model, we
plot the mass vs.\ radius relationship both with and without the inclusion of the PBS crust equation of state
\cite{PBS}.

Skyrme models employed in our work are softer at lower densities and stiffer at higher densities, 
compared to the relativistic mean
field model Z271v6. This results in generally smaller radii at a given mass for the Skyrme interactions
and a larger maximum mass. For neutron star masses greater than about 1.3$M_\odot$, all equations of
state give mass-radius relations consistent with the empirical constraints from Ref.\ \cite{Steiner2012}. 
The inclusion of a crust increases the radius by about 0.5\,km for a 1.4 $M_\odot$ 
neutron star and for the most massive neutron stars has a relatively small impact. This is, perhaps, a first indication that 
the crust equation of state used should be consistent with the underlying microphysics model given its impact in the size of typical neutron stars
(of mass 1.4 $M_\odot$). This and other implications of the crust EoS are current subject of investigation to be published in future works.

We see that none of the models employed reaches the maximal mass value of 2 solar masses,
indicating that the phenomenological interactions must be supplemented by
a stiffer EoS at high densities. We identify the density above which the equation of state must be
modified by employing the constant speed of sound parametrization \cite{alford15} at the superluminal
boundary. As a representative example, the Ska35s20.09 Skyrme interaction can generate a 2 
solar-mass neutron star if the equation of state beyond $\rho = 4.5\rho_0$ is taken to be that of a 
liquid with constant speed of sound equal to the speed of light. The presence of the crust has only a
very small effect, increasing the transition density to $\rho = 4.6\rho_0$. 

However, for the high densities required for 2 solar mass neutron stars there are no other constraitns on the properties of bayonic matter and our criteria is valid only for low desnities.
Thus, we would only like to emphasize that is feasible to reach the mass limit by modifications of the high density regime, but the proper study
of matter in this range is beyond the scope of this work. 

\section{Conclusion}
\label{summary}

We have studied the use of mean field models to reproduce zero- and finite-temperature 
nuclear equations of state derived from microscopic many-body theory with realistic chiral two- and 
three-body forces. Comparing to quantum Monte Carlo simulations employing two-body forces
alone, we find that the zero-temperature neutron matter equation of state is well converged in
perturbation theory up to half saturation density, with uncertainties on the order of a few percent. 
From this density regime we select mean field models consistent with the predictions from chiral
effective field theory as well as the available empirical infinite matter constraints. We then explore 
consistency with chiral nuclear thermodynamics and find that the free energy as a 
function of the temperature for mean field models are strongly correlated with the nucleon effective 
mass, with smaller values giving rise 
to larger kinetic energy contributions and smaller entropy contributions to the free energy.

In addition we studied the effects of the single-particle dispersion relation on select astrophysical
phenomena, such as neutrino absorption in the proto-neutron star neutrinosphere and the 
adiabatic evolution of core collapse. While the single-particle energy shift is most relevant for 
determining the charged-current weak reaction rates, the nucleon effective mass largely governs 
the isentropic temperature-density relation. For the latter quantity, significant variations are observed 
in the neutron matter low-entropy 
regime. At the largest densities considered ($\rho \sim 1.6 \rho_0$), the temperature can vary by up 
to 50\% for matter with an entropy per baryon of $S/N = 1$.

Finally, the mean field models consistent with the low-density chiral n3lo414 equation of state 
were used to explore the high-density regime relevant for cold neutron star matter where chiral
effective field theory is expected to break down. While the 
mass-radius curves are consistent with present observational constraints around $M\sim 1.5 M_\odot$,
all maximum neutron star masses lie below $M\sim 2 M_\odot$ and the constant speed of
sound parametrization was used to complete the mean field models in the high-density regime
above $\rho = 4.5 \rho_0$. 

The present work lays the foundation for future efforts to construct consistent equations of state and 
neutrino response functions that are compatible with chiral effective field theory for use in numerical
simulations of core-collapse supernovae and binary neutron star mergers.


\acknowledgments{We thank A.\ Bulgac and S.\ Reddy for useful discussions, 
A.\ Brown for sharing the parameters of the Skyrme models used in this work, and C.\ Wellenhofer
for providing us with the temperature- and density-dependent entropy from chiral nuclear forces. 
The work of J.\ W.\ Holt was supported by US DOE Grant No.\ DE-FG02-97ER41014. The work of A.\ Roggero
was supported by NSF Grant No.\ AST-1333607. 
The work of E.\ Rrapaj was supported by US DOE Grant No.\ DE-SC0008489.
Some of the most intensive computations have been performed at NERSC thank to a Startup allocation.}
 
 
\section{Appendix}
\label{appendix}

\subsection{Sign Problem}

\label{ssignproblem}

As explained in Section~\ref{smethod} the main systematic bias in the CIMC calculations 
come from imposing a fixed-node approximation in order to deal with the sign problem. In this section we 
discuss additional details and estimate the impact of this approximation. First we recall that within 
the fixed-node
approach one defines a family of {\it sign-problem-free} hamiltonians $H_{\gamma}$ 
(see e.g., Ref.\ \cite{tenHaaf}).
If we introduce the sign function
\begin{equation}
{\mathfrak s}({\bf m},{\bf n}) = sign\left( \frac{\langle \Phi_G\vert {\bf m}\rangle}{\langle {\bf n}\vert \Phi_G\rangle} \langle {\bf m}\vert H\vert {\bf n}\rangle\right),
\end{equation}
where $\Phi_G$ is the wavefunction used for the fixed node procedure, we can define the off 
diagonal matrix elements (${\bf n}\neq{\bf m}$) of the sign-problem-free Hamiltonian $H_{\gamma}$ as:
\begin{equation}
\langle {\bf m}\vert {\cal H_\gamma}\vert {\bf n}\rangle =
\left\{
\begin{array}{cc}
-\gamma\frac{\langle \Phi_G\vert {\bf m}\rangle}{\langle {\bf n}\vert \Phi_G\rangle} \langle {\bf m}\vert H\vert {\bf n}\rangle & {\mathfrak s}({\bf m},{\bf n}) > 0\\
\frac{\langle \Phi_G\vert {\bf m}\rangle}{\langle {\bf n}\vert \Phi_G\rangle} \langle {\bf m}\vert H\vert {\bf n}\rangle & {\rm otherwise}
\end{array}
\right. ,
\end{equation}
while the diagonal elements are:
\begin{equation}
\langle {\bf n}\vert {\cal H_\gamma}\vert {\bf n}\rangle =
\langle {\bf n}\vert H\vert {\bf n}\rangle + (1+\gamma)\sum_{
\substack{
{\bf n}\neq{\bf m}\\
{\mathfrak s}({\bf m},{\bf n})>0
}}
\frac{\langle \Phi_G\vert {\bf m}\rangle}{\langle {\bf n}\vert \Phi_G\rangle} \langle {\bf m}\vert H\vert {\bf n}\rangle
\end{equation}
It can be easily seen that for $\gamma = -1$ the original Hamiltonian is recovered. Furthermore if we label with $E_{\gamma}$
the lowest eigenvalue of a given $H_{\gamma}$ one can prove that any linear extrapolation from two values $E_{\gamma_1}$ and $E_{\gamma_2}$
at $\gamma_1,\gamma_2 \geq 0$ to $E^{\gamma_1,\gamma_2}_{-1}$ provides an upper bound on the true ground state energy $E_{-1}$.

In the calculations
presented here this procedure has been adopted by performing linear extrapolations using values at $\gamma=0$ and 
$\gamma=1$. The variation between the eigenvalues at different $\gamma$, including the extrapolated value at $\gamma=-1$
can be used as lower bounds on the missing energy contribution coming from the fixed-node approximation. We find
that this spread in the case of PNM is within the estimated error bars coming from statistical uncertainties and are 
usually less than 20\,keV per particle. In the case of SNM, however, the values at different gamma are 
outside the statistical
error bars for densities below $0.1-0.14$\,fm$^{-3}$ but still of the order of $100$\,keV per particle even at the 
lowest densities. 

This lower bound on the fixed-node error would be a good estimate provided there are no
relevant contributions beyond the linear one in the $\gamma$ extrapolation, a condition that cannot be checked 
in practice.
In order to have another quantitative estimate of the systematic bias introduced by using the fixed-node approximation,
we can check the sensitivity to changes in the underlying guiding wavefunction. This is analogous to the error estimates
in MBPT obtained by changing the single-particle spectrum. The guiding wavefunction used in our calculations comes
from the coupled-cluster double approximation with amplitudes obtained from second-order MBPT 
(see Ref.\ \cite{Roggero2013} for additional details) which clearly has
zero overlap with states composed of an odd number of particle-hole excitations. In order to check the sensitivity 
to this choice we can set a lower bound on the overlaps of this wavefunction to any state in the Hilbert space generated
by the single-particle space we have chosen. In this way we are including effectively a crude ansatz for triple and higher
cluster excitations in this wavefunction that we will call $CCD^{*}$.
\begin{figure}[t]
\begin{center}
\includegraphics[scale=0.33]{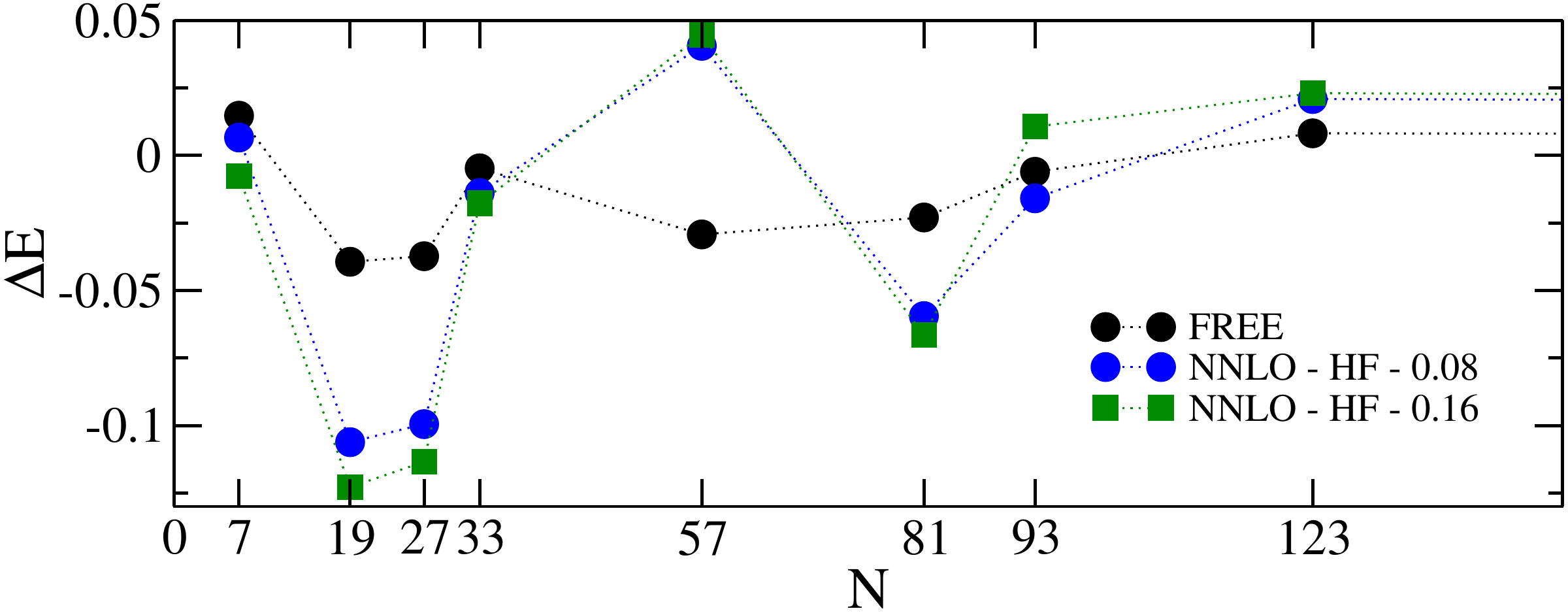}
\caption{(color online) Finite-size errors in the energy per particle as a function of the number of particles per 
spin-isospin species. The results shown are for a free gas as well as for Hartree-Fock 
calculations with the NNLO$_{opt}$ interaction at the two densities $\rho = \rho_0$ and $0.5\rho_0$. }
\label{Fig_fs} 
\end{center}
\end{figure}

\begin{figure*}[tbh]
\begin{center}
\includegraphics[scale=0.5]{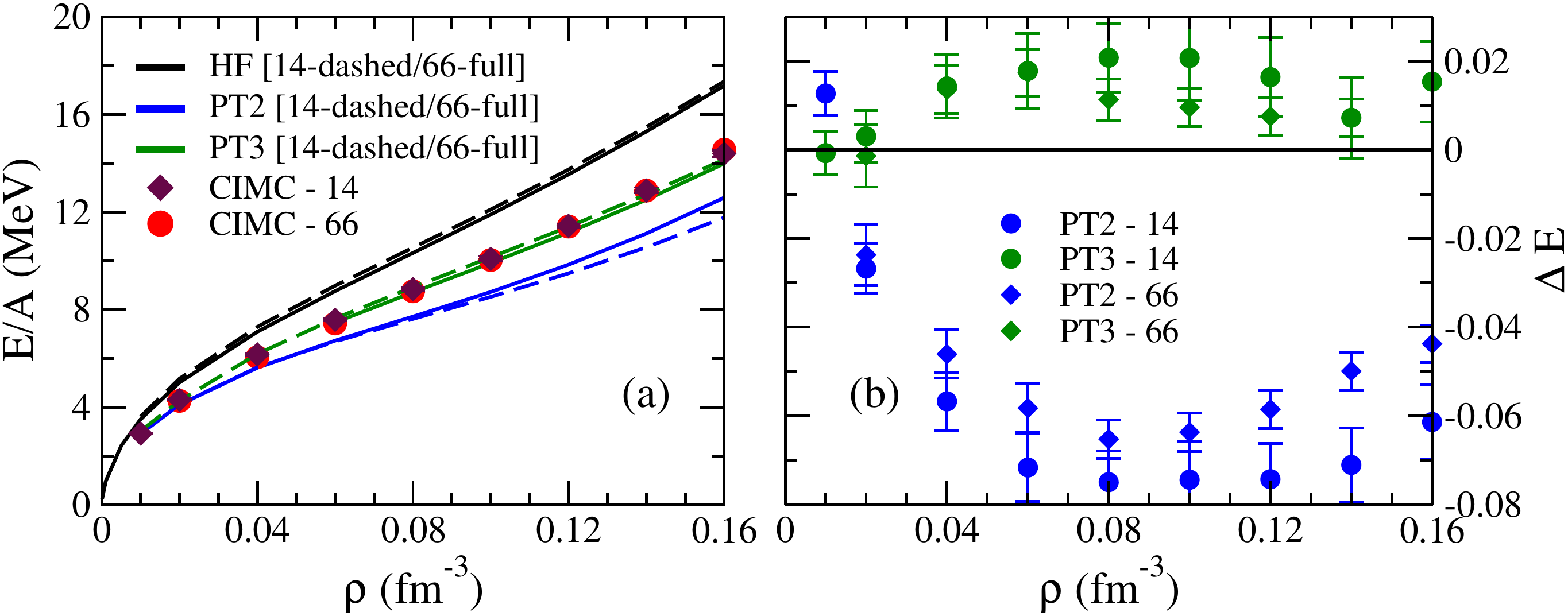}
\caption{ (color online) (panel (a)): energy per particle in PNM computed with $N = 14$ and $66$ neutrons from 
CIMC and MBPT using the NNLO$_{opt}$ chiral two-nucleon potential. (panel (b)): differences between the results
from CIMC and MBPT at second and third order using $N=14$ and $66$.
\label{Fig_nmat_fs}} 
\end{center}
\end{figure*}

The effect of using this as a guiding wavefunction 
in the case of PNM in the density range explored in this work is just to raise the statistical error on the 
ground state energy
and no difference can be seen outside the error bars, which suggests that the $CCD^{*}$ wavefunction provides
a worse nodal constraint (higher statistical error) but the bias introduced by neglecting triples and higher correlations
is smaller or comparable to the statistical errors in the calculation. This is also compatible with the estimates from 
coupled-cluster calculations using perturbative triples \cite{Hagen2014}. In the case of SNM the effect of
using the $CCD^{*}$ wavefunction is much stronger: the statistical error is comparable to the one obtained using 
the bare $CCD$ wavefunction but now the extrapolated eigenvalue $E_{-1}$ shows considerably less binding, and 
this effect is stronger as the density is lowered (indicating that the origin is probably the appearance of clustering in 
the system). For instance, in the case
of the \nnnlo~$414$ interaction we find changes of about 1\,MeV at $\rho = 0.02$\,fm$^{-3}$ and 
about 0.5\,MeV at $\rho=0.08$\,fm$^{-3}$.
Since both wavefunctions provide upper bounds on the energy, estimates are always obtained with the $CCD$
guiding wavefunction, but we can use the $CCD^{*}$ results to obtain a qualitative error band. 
This is the procedure we employ
to produce the grey band in Fig.\ \ref{Fig_smat_evscimc}.
\subsection{Finite Size Effects}
\label{fsize}

The quantum Monte Carlo calculations presented in Section \ref{benchmark} have been carried out using 
finite systems containing a fixed number of particles $N$. In order to reach the thermodynamic limit this 
number has to be taken as 
large as possible for a given density; however due to the strong dependence ($N^2$ for CIMC and 
PT2 and $N^3$ for part of PT3) of the computational time on the number of particles, all the 
results presented so far employed $N_s = 7$ for each spin/isospin species. Here we show that this
is sufficient for the low densities considered in Section \ref{benchmark}.

In order to avoid degenerate 
ground states we choose $N_s$ such that all states up to a Fermi
momentum $k_F$ have been filled, and in our cubic lattice this restricts the allowed particle numbers to
take on the values shown on the $x$-axis
in Fig.\ \ref{Fig_fs}. Among these {\it closed-shell} configurations the ones that preserve the cubic symmetry of
the underling lattice (corresponding to $N_s = 7,33,\dots$) 
have smaller deviations from the $N_s \to \infty$ limit 
as can be seen from Fig.\ \ref{Fig_fs} and are usually selected so as to minimize finite size effects. 
It is remarkable
that for a small system with $N_s = 7$ both the free-particle energy and the interacting Hartree-Fock 
energies are converged to the thermodynamic limit at the $\approx 1-2\%$ level, similar to the 
$N_s = 33$ system usually employed in
similar calculations. 

In the left panel of Fig.\ \ref{Fig_nmat_fs} we plot the energy per particle in PNM 
for two systems composed of either $N=14$ or $N=66$ neutrons interacting through the NNLO$_{opt}$ 
interaction obtained both with CIMC and  
with different orders of MBPT. The differences are very small for all calculations apart from second-order
perturbation theory (PT2), which for $N=66$ shows a stronger density dependence above 
$\rho \approx 0.10$\,fm$^{-3}$. The situation is similar
with the other interactions used throughout this work. Since our main focus is to explore the convergence 
of MBPT calculations we plot in the right panel of Fig.\ \ref{Fig_nmat_fs} the relative difference between 
CIMC and both PT2 and PT3 for the case of $N=14$ particles (black and red dots respectively) and $N=66$ 
particles (green and blue dots). We see that the 
conclusions on the expected convergence errors are consistent for densities $\rho \lesssim 0.10$\,fm$^{-3}$, 
a density at which
three-body interactions are already important. We therefore conclude that our $N_s=7$ particle systems are
large enough for our purposes in the density region where two-body interactions are the dominant contribution.
\begin{table*}
\begin{center}
\begin{tabular}{c|ccc|ccc}
     \hline\hline
\noalign{\smallskip}
PNM   &  & NNLO$_{opt}$ &  &  & \nnnlo$-414$& \\
\hline 
\noalign{\smallskip}
$\rho\; (fm^{-3})$  & $\Delta$PT2$_{free}$ & $\Delta$PT2$_{hf}$ & $\Delta$(PT2$_{free}$-PT2$_{hf}$) &  $\Delta$PT2$_{free}$ & $\Delta$PT2$_{hf}$ & $\Delta$(PT2$_{free}$-PT2$_{hf}$)   \\
\hline 
\noalign{\smallskip}
0.02  &  -0.0422(55)  & -0.0268(56) & -0.015449(90) & -0.0239(46)  & -0.0131(47) & -0.010774(51) \\
0.04  &  -0.0863(64)  & -0.0567(66) & -0.02961(21)  & -0.0408(48)  & -0.0253(48) & -0.015511(77) \\
0.06  &  -0.1175(73)  & -0.0716(76) & -0.04590(38)  & -0.0473(51)  & -0.0287(52) & -0.018634(99) \\
0.08  &  -0.1381(71)  & -0.0749(71) & -0.06320(48)  & -0.0490(52)  & -0.0282(54) & -0.02079(11) \\
\noalign{\smallskip}
     \hline\hline
\noalign{\smallskip}
SNM   &  & NNLO$_{opt}$ &  &  & \nnnlo$-414$& \\
\hline 
\noalign{\smallskip}
$\rho\; (fm^{-3})$  & $\Delta$PT2$_{free}$ & $\Delta$PT2$_{hf}$ & $\Delta$(PT2$_{free}$-PT2$_{hf}$) &  $\Delta$PT2$_{free}$ & $\Delta$PT2$_{hf}$ & $\Delta$(PT2$_{free}$-PT2$_{hf}$)   \\
\hline 
\noalign{\smallskip}
0.04  &  0.1711(15)  & 0.0217(13)  & 0.14940(19)  & 0.166(12)   & 0.040(11)  & 0.1264(14) \\
0.06  &  0.2829(12)  & 0.0946(10)  & 0.18832(17)  & 0.1965(67)  & 0.0540(59) & 0.14250(80) \\
0.08  &  0.31699(81) & 0.10439(68) & 0.21260(13)  & 0.2185(81)  & 0.0662(71) & 0.1523(10) \\

\noalign{\smallskip}\hline\hline
\end{tabular}
\end{center}
\caption{Deviations from CIMC energies of $2$nd order MBPT calculations for both PNM (top part) and SNM (bottom part)
for two of the interactions used in this work. $\Delta$PT2$_{free}$ is the relative difference with respect to
MBPT2 with a free spectrum in the propagators, $\Delta$PT2$_{hf}$ instead uses a Hartree-Fock dispersion while
$\Delta$(PT2$_{free}$-PT2$_{hf}$) simply states the variation coming from changes in the single-particle spectrum. Errors
in parenthesis are statistical errors coming from the CIMC result.\label{Tab_hfvsfree}}
\end{table*}

\subsection{Effects of the Single Particle Spectrum} 
\label{sspspectrum}
For all calculations presented the Hartree-Fock spectrum has been used in the single particle propagators (see e.g., 
Ref.\ \cite{ShavittBartlett} for details). We find that this choice yields a considerable improvement in the convergence
pattern of the perturbative calculations. As can be seen from the results in Table \ref{Tab_hfvsfree} for instance,
the errors at second order are approximately twice as large using a free spectrum as the ones employing
self-energy corrections at the Hartree-Fock level. For nuclear matter this factor can be even larger,
and it does not seem connected to the perturbativeness of the interaction as can be deduced from the increased
difference in going from the harder NNLO$_{opt}$ to the softer \nnnlo~414 potential.

We conclude by pointing out
that using the difference in energies obtained by varying the single-particle dispersion relation gives a
qualitative understanding of the convergence of the perturbative
calculations, with \nnnlo~414 variations being smaller than the NNLO$_{opt}$ ones and with a substantial
increase in going from PNM to SNM. It has to be noted however that these variations cannot be used to give 
a quantitative estimate of the errors in the many-body calculation, which is evident from the SNM results
where $\Delta$(PT2$_{free}$-PT2$_{hf}$) overestimates the error by a factor of $\approx 2-3$. This is however
not conclusive since in SNM the CIMC calculation is not fully under control. In PNM where the 
difference with CIMC is a more reliable check of convergence, the errors are consistently underestimated by
employing $\Delta$(PT2$_{free}$-PT2$_{hf}$).

\bibliography{biblio}

\end{document}